\documentclass{aa}  
\usepackage{array}
\usepackage{pifont}
\usepackage{tabularx}

\usepackage{graphicx}
\usepackage{txfonts}
\usepackage{multirow}
\usepackage[version=4]{mhchem}
\usepackage[colorlinks=true,
            linkcolor=red,
            urlcolor=blue,
        citecolor=blue]{hyperref}
\usepackage[dvipsnames]{xcolor}
\usepackage{placeins}
\usepackage{caption}
\usepackage{subcaption}

\newcommand{\R}[0]{\ensuremath{R}}
\newcommand{\Rv}[1]{\ensuremath{R = #1}}
\newcommand{\SN}[0]{\ensuremath{S/N}}
\newcommand{\SNv}[1]{\ensuremath{S/N = #1}}

\newcommand{\mic}[1]{\ensuremath{#1}~\textmu m}

\newcommand{\pt}[0]{\textit{P}$-$\textit{T}}

\newcommand{\Rpl}[0]{\ensuremath{R_{\text{pl}}}}
\newcommand{\Mpl}[0]{\ensuremath{M_{\text{pl}}}}

\newcommand{\Ps}[0]{\ensuremath{P_0}}
\newcommand{\Ts}[0]{\ensuremath{T_0}}

\newcommand{\life}[0]{LIFE}
\newcommand{\hwo}[0]{HWO}
\newcommand{\hwolife}[0]{HWO+LIFE}
\newcommand{\uvvisnir}[0]{UV/VIS/NIR}
\newcommand{\mir}[0]{MIR}
\newcommand{\lifesim}[0]{LIFE\textsc{sim}}
\newcommand{\prt}[0]{\texttt{petitRADTRANS}}
\newcommand{\pI}[0]{LIFE Paper~I}
\newcommand{\pII}[0]{LIFE Paper~II}
\newcommand{\pIII}[0]{LIFE Paper~III}
\newcommand{\pV}[0]{LIFE Paper~V}
\newcommand{\pIIIaV}[0]{LIFE Papers~III and~V}

\newcommand\crule[3][black]{\textcolor{#1}{\rule{#2}{#3}}}
\definecolor{blue}{RGB}{100, 143, 255}
\definecolor{magenta}{RGB}{220, 38, 127}
\definecolor{gold}{RGB}{255, 176, 0}

\begin{document}

     \title{Large Interferometer For Exoplanets (LIFE):}
   
  % \subtitle{N-1. Synergies between the mid-infrared space interferometer \life~and  a \uvvisnir~Habitable Worlds Observatory (HWO) concept}
\subtitle{XIII. The Value of Combining Thermal Emission and Reflected Light for the Characterization of Earth Twins}
   \titlerunning{LIFE: XIII.  The Value of Combining Thermal Emission and Reflected Light for the Characterization of Earth Twins}
    \authorrunning{Alei et al.}
   \author{E. Alei
          \inst{1,2,3}, S.~P. Quanz\inst{1,2,4}, B.~S. Konrad\inst{1,2}, E.~O. Garvin \inst{1},  V. Kofman\inst{5,6}, A. Mandell\inst{5}, D. Angerhausen\inst{1,2}, P. Molli\`ere\inst{7}, M.~R. Meyer \inst{8}, T. Robinson\inst{9}, S. Rugheimer\inst{10}
    and the \textit{LIFE} Collaboration\thanks{Webpage: \url{www.life-space-mission.com}}
          }

   \institute{ ETH Zurich, Institute for Particle Physics \& Astrophysics, Wolfgang-Pauli-Str. 27, 8093 Zurich, Switzerland\\
    e-mail: \texttt{elalei@phys.ethz.ch; eleonora.alei@nasa.gov} 
    \and
    National Center of Competence in Research PlanetS (www.nccr-planets.ch)
    \and NASA Postdoctoral Program Fellow, NASA Goddard Space Flight Center, Greenbelt, MD, USA
    \and ETH Zurich, Department of Earth Sciences, Sonneggstrasse 5, 8092 Zurich, Switzerland
    \and  
    NASA Goddard Space Flight Center, Greenbelt, MD, USA
    \and 
    American University, 4400 Massachusetts Ave NW, Washington, DC 20016
    \and Max-Planck-Institut f\"ur Astronomie, Königstuhl 17, 69117 Heidelberg, Germany
    \and Department of Astronomy, The University of Michigan West Hall 323, 1085 S. University Avenue Ann Arbor, MI 48109
    \and Lunar \& Planetary Laboratory, University of Arizona, Tucson, AZ, USA
        \and Department of Physics and Astronomy, York University, 4700 Keele St., Toronto, ON M3J 1P3, Canada
             }

   \date{Received ; accepted }

% \abstract{}{}{}{}{} 
% 5 {} token are mandatory

  \abstract
  % context heading (optional)
   {Following the recommendations to NASA (in the Astro2020 Decadal survey) and ESA (through the Voyage2050 process), the search for life on exoplanets will be a priority in the next decades. Two concepts for direct imaging space missions are being developed for this purpose: the Habitable Worlds Observatory (HWO), and the Large Interferometer for Exoplanets (LIFE). These two concepts operate in different spectral regimes: HWO is focused on reflected light spectra in the ultraviolet/visible/near-infrared (\uvvisnir), while LIFE will operate in the mid-infrared (\mir) to capture the thermal emission of temperate exoplanets.}
  % aims heading (mandatory)
   {In this study we aim to assess the potential of HWO and LIFE in characterizing a cloud-free Earth twin orbiting a Sun-like star at 10 parsec distance both as separate missions and in synergy with each other. We aim to quantify the increase in information that can be gathered by joint atmospheric retrievals on a habitable planet.}
  % methods heading (mandatory)
   {We perform Bayesian retrievals on simulated data obtained by a \hwo-like and a \life-like mission separately, then jointly. We consider the baseline spectral resolutions currently assumed for these concepts and use two increasingly complex noise simulations, obtained using state-of-the-art noise simulators.  }
  % results heading (optional), leave it empty if necessary 
   {A \hwo-like concept would allow to strongly constrain \ce{H2O}, \ce{O2}, and \ce{O3}, in the atmosphere of a cloud-free Earth twin, while the atmospheric temperature profile is not well constrained (with an average uncertainty $\approx$ 100 K). \life-like observations would strongly constrain \ce{CO2}, \ce{H2O}, \ce{O3} and provide stronger constraints on the thermal atmospheric structure and surface temperature (down to $\approx$ 10 K uncertainty). For all the investigated scenarios, both missions would provide an upper limit on \ce{CH4}. A joint retrieval on \hwo~and \life~data would accurately define the atmospheric thermal profile and planetary parameters. It would decisively constrain \ce{CO2}, \ce{H2O}, \ce{O2}, and \ce{O3} and find weak constraints on \ce{CO} and \ce{CH4}. The significance of the detection is in all cases greater or equal than the single-instrument retrievals.}
    {Both missions provide specific information that is relevant for the characterization of a terrestrial habitable exoplanet, but the scientific yield can be maximized by considering synergistic studies of \uvvisnir+\mir~observations. The use of \hwo~and \life~together will provide stronger constraints on biosignatures and life indicators, with the potential of being transformative for the search for life in the universe.}

   \maketitle
%
%-------------------------------------------------------------------

\section{Introduction}

The characterization of temperate terrestrial exoplanets and the search for signatures of biology on these worlds are two primary goals that the astronomical community will strive to achieve in the next decades.  In general, current and near-future observatories will not be able to directly detect the signal of the atmospheres of habitable terrestrial planets: their flux is too small compared to their host stars, and their small angular separation poses a challenge for current high-contrast imaging instrumentation. Only some specific targets will be available for near-term studies (e.g., planets orbiting nearby stars, or systems orbiting M dwarfs) through transit spectroscopy (using, e.g., JWST, see  \citet{2016ApJ...817...17G,2024A&A...683A.212D}) or direct imaging (using e.g., ELT/METIS, see \citet{2015IJAsB..14..279Q,2021A&A...653A...8B}). Nevertheless, none of the currently planned missions will provide direct measurements of the atmospheric composition of a statistically significant sample of terrestrial, temperate exoplanets around Sun-like stars. We will require more sensitive observatories and instruments to directly image a sample of rocky temperate exoplanets and determine what are considered the prime targets for the search for life beyond the Solar System. 

The US Astro2020 Decadal survey \citep{NAP26141} recommended the pursuit of a technical and scientific study for the Habitable Worlds Observatory (HWO), an ultraviolet/visible/near-infrared (\uvvisnir) “high-contrast direct imaging mission with a target off-axis inscribed diameter of approximately 6 meters”, which shares design and technology heritage with the pre-Decadal mission concept studies HabEx \citep[Habitable Exoplanet Observatory,][]{Gaudi2020}  and LUVOIR \citep[Large UV/Optical/IR Surveyor,][]{Peterson2017}. At the same time, the ESA Voyage 2050 Senior Committee report \citep{Voyage2050} recommended the study of temperate exoplanets in the mid-infrared (\mir) as a potential strategy for the upcoming decade. A European-led effort to develop the \mir~space-based nulling interferometer LIFE \citep[Large Interferometer For Exoplanets,][]{2022ExA....54.1197Q,LIFEI} is currently being pursued for this purpose. In virtue of their different starlight suppression technologies, these two facilities will likely cover partly different regions of the target parameter space. LIFE will offer more flexibility to detect and characterize planets orbiting F, G, K, and M stars, thanks to the large effective aperture size enabled by ultra-stable nulling interferometry.  HWO will be focusing on F, G, and K stars since the inner working angle of a coronagraph impacts the detection of habitable zones closer to the star (e.g., Earth-like planets orbiting late K and M dwarfs). Still, there is an overlap in planets that will be studied by both missions -- even in the realm of already-known exoplanets, as shown in \citet{CarrionGonzalez2023}.

Both concepts are still at the early stages of their development, and we therefore rely on notional designs and approximations to better define the scope and the current technical challenges related to both missions. 
The community has been implementing strategies to simulate observations and data analyses of specific planet archetypes to gather information on the minimum and preferred mission requirements. 

In this context, Bayesian atmospheric retrieval frameworks have been and remain a valuable methodology \citep[see, e.g.,][]{2018haex.bookE.104M}. Such frameworks apply Bayes’ theorem to infer probability distributions of planetary properties from simulated empirical data and provide a statistically robust method to select the best theoretical model that best explains the data. This makes them not only the gold standard for analyzing and interpreting observations but also essential analysis tools when it comes to designing future missions. 
Simulated observations that take into account different architectures of the instruments can be fed to a Bayesian retrieval framework, to know what the retrieval process would infer from each hypothetical observation. This allows us to evaluate different architectures in terms of wavelength range, spectral resolution $R$, and signal-to-noise ratio \SN, as a function of specific mission goals.

Various studies of this kind have been carried out for generic/idealized versions of HabEx/LUVOIR-type instruments \citep[e.g.,][]{Feng_2018,DamianoHu2022,RobinsonSalvador2023,Latouf2023b, Latouf2023}, focusing on specific planet archetypes and simplifying assumptions about bandpass limitations and noise simulation. For HWO, a similar approach is being used to define a more accurate observing strategy for various planetary scenarios \citep[e.g.,][]{Young2023}.

In a previous study assessing the capabilities of the LIFE mission concept by \citet{Konrad2022} (\pIII), we performed atmospheric retrievals on a cloud-free Earth twin at a 10 pc distance to understand the minimum requirements to correctly characterize biosignatures of a living planet. We expanded our work in \citet{Alei2022} (\pV), where we analyzed the potential of a LIFE-like mission to characterize the Earth at various stages of its bio-geological evolution. Further studies have also been performed on a Venus twin \citep[i.e., LIFE Paper IX]{Konrad2023} and on real Earth satellite data \citep{Mettler2023}. Exploratory studies on the detectability of phosphine and capstone signatures have also been carried out within the LIFE series \citep[i.e., LIFE Papers VIII and XII]{2023AsBio..23..183A,Angerhausen_2024}.

When it comes to characterization, the study of exoplanetary atmospheres at various wavelengths provides us with separate unique windows into the physics and chemistry of the target.
From the chemical point of view, specific molecules can be more spectrally active in one wavelength range and only be constrained through observations at those characteristic wavelengths (e.g., \ce{O2} in the \uvvisnir~and \ce{CO2} in the \mir).
When it comes to physical quantities and processes, reflected light observations could provide us with knowledge of the dynamics, cloud composition, and albedo of the planet, as well as having direct access to surface biosignatures and ocean glint \citep[see, e.g.,][]{WILLIAMS2008927,Sagan1993}. On the other hand, thermal emission observations would provide information on the planetary dimensions, the thermal structure of the atmosphere, and its composition by directly measuring the planetary radiation, without having to rely on the knowledge of the stellar spectrum.

Observations in each wavelength range come with their drawbacks and degeneracies. In reflected light, radius and albedo are extremely coupled, especially in the case where clouds are present \citep[see, e.g.,][]{Feng_2018}. Yet, while complicated, observations in this wavelength range could still infer some information on hazes and cloud production in planetary atmospheres \citep[see, e.g., ][]{RobinsonSalvador2023}, which is essential for an accurate energy balance calculation and for ultimately understanding the nature of a given atmosphere. On the other hand, observations in the \mir~would be less sensitive to patchy clouds (see, e.g., \pV), allow for higher confidence in the characterization of the thermal and chemical composition of the atmosphere, and provide us with a more precise measurement of the radius (e.g., from the search phase of LIFE, see  \citet{LIFEII} i.e., \pII) which would help to disentangle the radius-albedo degeneracy. 

In this work, we provide a first qualitative assessment of the scientific impact of a synergistic approach between the two potential future missions. We further provide an overview of the unique strengths of missions exploring the \uvvisnir~and \mir~wavelength range to characterize a cloud-free Earth-like planet.

We aim to answer the following research questions: 1) What is the scientific potential of these two separate missions that explore two different wavelength regions? 2) To what extent would a joint atmospheric retrieval of \uvvisnir~and \mir~data improve the characterization of a cloud-free Earth twin? Would the detection of the most relevant biosignatures and habitability-related species improve by considering data from both wavelength ranges? 3) How would the assumed noise model affect the results when analyzing data from different instruments separately and jointly?

These are urgent questions to answer to provide realistic mission requirements, inform the design phase, and plan precursor studies accordingly. Such a study is critical at this point in time, as the community working on future-generation missions is being formed and organized. Understanding synergies between US-led and Europe-led missions is relevant to motivating and uniting the community in supporting both.

In the remainder of the manuscript, we approximate the HWO concept with a 6-meter inscribed LUVOIR-B-like instrument, to be able to leverage the knowledge acquired and the tools developed for the LUVOIR Final Report \citep{2019arXiv191206219T}, which are yet to be modified in any significant way for HWO. For the \life~concept, we consider the currently-assumed baseline architecture \citep[][i.e., Papers I and II of the LIFE Series]{LIFEI,LIFEII}.

 We describe how we produced the simulated observations and the updates we made on the atmospheric retrieval framework in Section \ref{sec:methods}. We describe in Section \ref{sec:results} our results for the sets of retrievals, considering both single-instrument and joint retrievals in each set. We compare the single-instrument results with previous studies and we discuss the impact of multi-instrument observations, as well as our limitations in Section \ref{sec:discussion}. We draw our conclusions in Section \ref{sec:conclusions}.

%--------------------------------------------------------------------

\section{Methods}\label{sec:methods}

In this section, we describe the calculation of the input spectrum and the simulated observed spectra, updates to the Bayesian retrieval framework compared to previous studies (see \pIIIaV), and the details of the sets of retrievals we consider in this study.

\subsection{Input spectrum}\label{sec:inputspectrum}

For all retrievals performed in this study, we assumed a simplified cloud-free Earth-like atmosphere. We assumed a pressure-independent atmospheric composition (i.e., the composition was the same for all atmospheric layers) and the pressure-temperature (\pt) profile was assumed to be a fourth-order polynomial (5 parameters) that represented the best fit to the U.S. Standard Atmosphere 1976 \citep{united1976u}. The use of such simplified atmospheric profiles allowed us to reduce the uncertainty linked to the variability of the atmospheric profiles and abundances. The input values for each parameter used in the retrieval are shown in Table \ref{table:parameters}. 

Assuming this simplified atmosphere, we produced a theoretical spectrum using \prt~\citep{Molliere2019}. To calculate the spectrum, we used HITRAN 2020 opacity tables, calculated assuming air broadening and a line cutoff of 25 cm$^{-1}$ (see Table \ref{table:opacities}) and UV opacities for \ce{CO2}, \ce{O2}, \ce{H2O}, \ce{O3}, \ce{CH4}, \ce{N2O}, and \ce{CO} from the MPI-MAINZ UV/VIS Spectral Atlas \citep{essd-5-365-2013}. We also considered all possible collision-induced absorption (CIA) opacities and Rayleigh scatterers (see Table \ref{table:cia}).

The star-planet system is assumed to be at a 10 pc distance. This distance has been used in previous studies (see \pIIIaV). The planet was assumed to be illuminated on only one hemisphere by a Sun-like star ($T_\star=5778$~K, $R_\star=~1~\mathrm{R_\odot}$) as if the system was face-on, or an edge-on planetary orbit seen at quadrature. To do this, we employ the scattering of direct light treatment implemented in \prt, which is explained in more detail in Appendix \ref{app:scattering}.

\subsection{Simulating observed spectra} \label{sec:observedspectra}

In this work, we simulate three sets of observations assuming 1) a high-resolution low-noise case (used for validation), 2) a baseline resolution and simplified noise scenario, and 3) a baseline resolution with a higher-fidelity noise calculation. 
 For each set, we produce two simulated observed spectra (one for the \uvvisnir~and one for the \mir~wavelength ranges respectively), assuming different spectral resolutions and noise instances.

For the validation set (see Appendix \ref{sec:photonnoise}), we assume a spectral resolution of \Rv{1000} and we only consider photon noise, whose \SN~ is 50 at a reference wavelength point (\mic{0.55} for \hwo~and \mic{11.2} for \life). Such a high-resolution low-noise scenario was chosen to both validate the retrieval framework after the updates that were performed, as well as to simulate the proof of concept of what could be the performance of such idealized missions. The choice of the reference wavelength points was based on the continuum emission of the spectrum and it has been previously adopted by other studies, such as \citet{Feng_2018} for LUVOIR and \citet{Konrad2022} and \citet{Alei2022} for LIFE (\pIIIaV).

\begin{figure}[!ht]
    \centering
    \includegraphics[width=\linewidth]{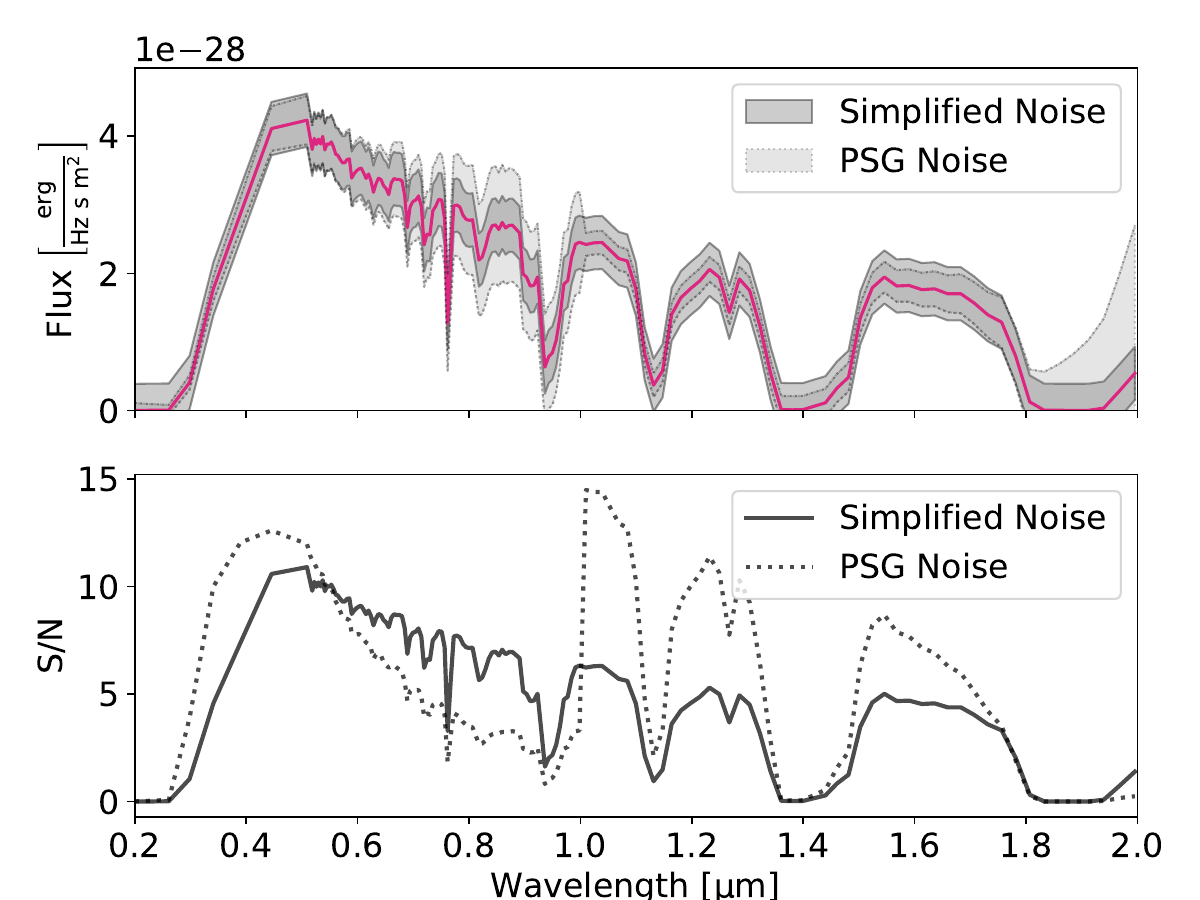}
    \caption{Simulated observation of an Earth twin by the \hwo~concept assuming a simplified noise and a higher-fidelity noise instances. \emph{Top panel}: The simulated observed flux at 10 pc distance between 0.2 and \mic{2} for the two noise scenarios: simplified noise at constant error bars (dark gray shaded area with solid edges) and higher-fidelity simulated noise obtained with PSG (light gray shaded area with dotted edges). \emph{Bottom panel}: Signal-to-noise ratio (\SN) as a function of wavelength for the two noise instances: simplified noise (solid line) and PSG-simulated noise (dotted line).  }
    \label{fig:luvoirinput}
\end{figure}

\begin{figure}[!ht]
    \centering
    \includegraphics[width=\linewidth]{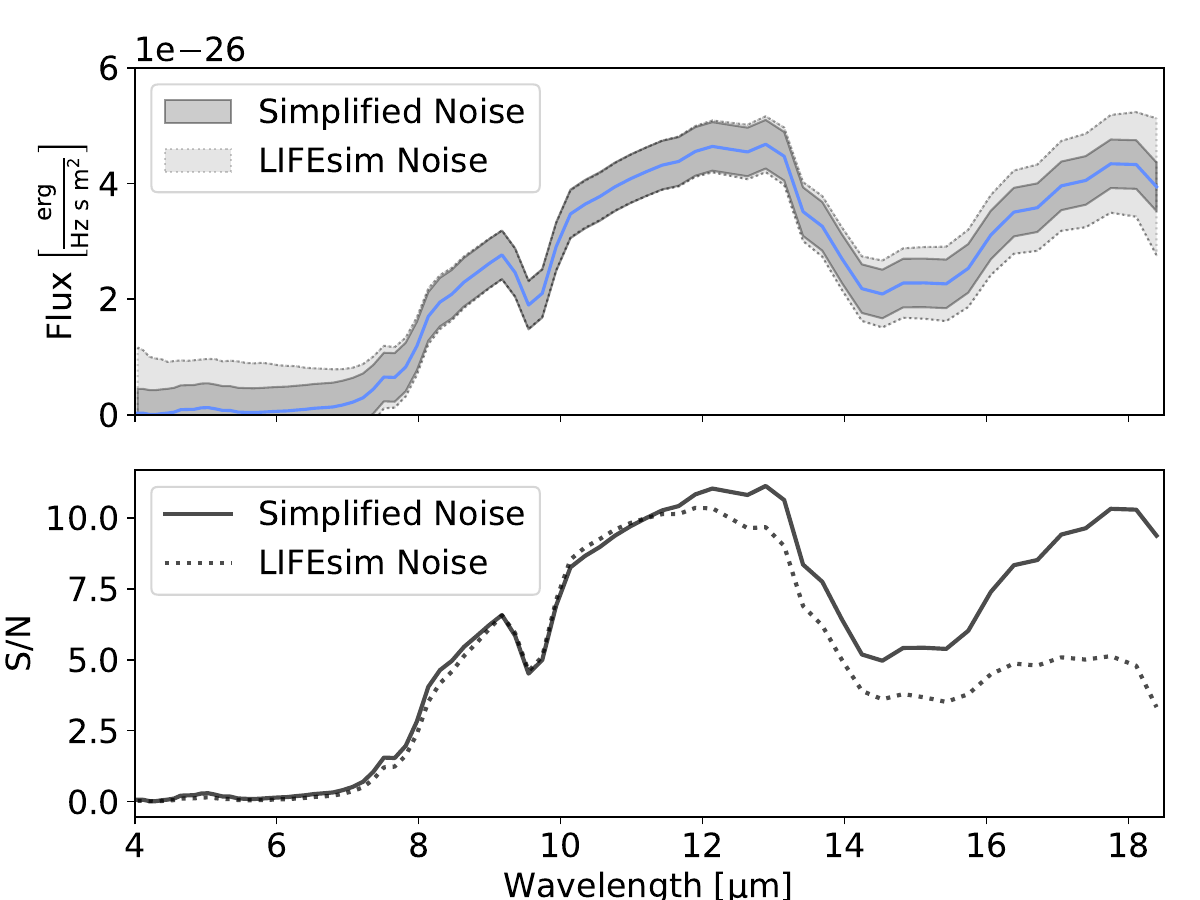}
    \caption{Simulated observation of an Earth twin by the \life~concept assuming a simplified noise and a higher-fidelity noise instances. \emph{Top panel}: The simulated observed flux at 10 pc distance between 4 and \mic{18.5} for the two noise scenarios: simplified noise at constant error bars (dark gray shaded area with solid edges) and higher-fidelity simulated noise obtained with \lifesim\ (light gray shaded area with dotted edges). \emph{Bottom panel}: Signal-to-noise ratio (\SN) as a function of wavelength for the two noise instances: simplified noise (solid line) and \lifesim-simulated noise (dotted line). }
    \label{fig:lifeinput}
\end{figure}

The other two sets of retrievals assume more realistic spectral resolutions and noise values, considering two different noise instances. The first set considers a more simplified noise scenario, while the second set assumes higher-fidelity noise simulations obtained with the available noise simulators for the two missions. The simulated observations and the corresponding wavelength-dependent \SN~for both missions are shown in Figures \ref{fig:luvoirinput} and \ref{fig:lifeinput}.


Concerning the simplified noise set, we assumed the current baseline resolutions for the instruments (\Rv{7} between \mic{0.2-0.51}, \Rv{140} between \mic{0.51-1}, \Rv{70} between \mic{1-2.0}  for \hwo~considering the values from the LUVOIR-B concept; \Rv{50} between \mic{4-18.5} for LIFE). We considered constant error bars (i.e., the same flux value for uncertainty)  at each flux point. The value of the flux uncertainty was calculated to correspond to a value of \SNv{10} at the two reference wavelength points. Then, that constant value was assumed throughout the whole wavelength range for the two instruments. This implementation has been used in the literature for \uvvisnir~studies 
\citep[see, e.g., ][]{Feng_2018, DamianoHu2022, RobinsonSalvador2023,Young2023}.

In the higher-fidelity simulated noise set, we assumed more realistic uncertainty values using state-of-the-art noise simulators for the two concepts. 
The theoretical input spectrum was fed to both the \life~simulator \lifesim\ \citep[as described in][i.e., \pII]{LIFEII} and the \hwo~noise simulator included in the Planetary Spectrum Generator\footnote{\url{https://psg.gsfc.nasa.gov/index.php}} \citep{2022fpsg.book.....V}. 
The settings used in \lifesim~to produce a simulated \life~observation are shown in Table \ref{tab:lifesim}.  The settings used in the PSG LUVOIR-B simulator are collected in Table \ref{tab:luvoirsim}.

We set up the simulators to be as similar as possible, though these are at a different state of their development. The \life~simulator \lifesim\ considers astrophysical noise, while the PSG \hwo~simulator includes also instrumental noise. However, given the outcome of the Decadal Survey, the \hwo~simulator will probably undergo significant changes to be adapted to the HWO concept. Still, these two simulators represent the current knowledge of the architecture and the expected noise from the two missions, which will be further refined in the upcoming years.

The S/N generated by \lifesim~is scaled to a reference value of 10 at \mic{11.2}, similar to the previous papers in the \life~series (\pIIIaV). For \hwo, we calculated the S/N for each of the wavelength portions of the spectrum separately, so that it would have a value of \SNv{10} at three reference wavelength points: \mic{0.35} for the UV range (\Rv{7} between \mic{0.2-0.51}); \mic{0.55} for the VIS range (\Rv{140} between \mic{0.51-1}); \mic{1.2} for the NIR range (\Rv{70} between \mic{1-2.0}). Also in this case, the choice of the wavelength reference points for the UV  and NIR was based on the peak of the continuum emission in the bands.

Compared to the simplified noise scenario, the \hwo~\SN~changes up to a factor of about two at wavelength bands that are featured by specific spectral signatures (see bottom panel of Figure \ref{fig:luvoirinput}): the noise around the UV \ce{O3} line between \mic{0.25-0.4}  and in the NIR range between \mic{1-1.5} is lower compared to the previous set of runs that considered a simplified noise scenario. For \life, the noise almost doubles at wavelengths between \mic{15-18.5} (see bottom panel of Figure \ref{fig:lifeinput}) compared to the simplified noise scenario.

For all retrievals, we consider the same physical parameters of the star-planet system (Table \ref{tab:physical}). These values are consistent with previous studies in the \life~series except for the exozodi level. While previous works in the series assumed a value of 3 times the zodiacal dust level (see, e.g., \pIIIaV), the PSG LUVOIR-B simulator assumed a default value of 4.5 times the local zodiacal dust level. We therefore assumed the higher value of the two to compare results.

In both instruments, the throughput acting on a point source is dependent on the source position relative to its host star and the wavelength of the observation. Therefore, the shape of the observed spectra will also depend on the planet's position. In this work, for the sake of correcting for this effect, we assume perfect knowledge of the position of the planet.
Hence, the observed flux and its related noise are normalized by the wavelength-dependent throughput of the system. 
We do not randomize each flux data point by the noise. Rather, we assume that the noise is the uncertainty on the theoretical data flux. We will briefly discuss the implications of this approximation in Section \ref{sec:limitations}.

\subsection{Updates on the Bayesian retrieval framework and setup}\label{sec:updates}

The Bayesian atmospheric retrieval framework used in previous studies (e.g., \pIIIaV) has been updated to allow the correct simultaneous retrieval of both \mir~and \uvvisnir~spectra. Generally, any Bayesian atmospheric retrieval framework is composed of two main modules. First, a parameter estimation module \citep[\texttt{pyMultiNest} in this case,][]{Buchner:PyMultinest} to iteratively sample the prior space to retrieve the best subset of parameters that explains the data (the posteriors). Second, a forward model \citep[\prt~in this case][]{Molliere2019} is used to calculate the spectrum corresponding to a set of parameters. The majority of the updates were performed to allow the use of features that were implemented in \prt~for \pV~\citep[see Appendix A of][]{Alei2022} but that were not considered in the previous studies.

Firstly, we adapted the existing retrieval framework to handle multiple input spectra simultaneously to account for different resolutions, signal-to-noise ratios, and wavelength ranges that are specific to the different instruments. We achieve this by calculating theoretical spectra at a resolution \Rv{200}, which is higher than the maximum foreseen resolution, and then binning these down to the input wavelength points at each retrieval iteration. For this step, we used the spectral rebinning tool \texttt{SpectRes} \citep{Carnall:SpectRes}.

Secondly,  we consider the atmospheric and surface scattering of both direct light and thermal atmospheric radiation as additional processes compared to previous studies. We assume to know the stellar radius and temperature, incident angle, and the distance of the stellar-planet system fixed in the retrievals. We now introduce the surface reflectivity $r_s$ as a parameter to be retrieved in the various runs. In the case of a thin, cloud-free atmosphere, the surface reflectance makes up for the majority of the albedo of the planet. Therefore, having $r_s$ as a free parameter allows us to qualitatively analyze any degeneracy between albedo and radius at UV-VIS wavelengths \citep[see, e.g.,][]{Gaudi2020,Feng_2018}. We refer to Appendix \ref{app:scattering} for more details on the scattering treatment and the definition of the surface reflectance.

In the retrieval, we consider the same set of opacities that were used to calculate the input spectrum (see Table \ref{table:opacities}). The retrieval framework also takes into account collision-induced absorption and Rayleigh scattering (see Table \ref{table:cia}). This choice was made to minimize the biases that the use of different opacity line lists might cause. We refer to \citet{AleiSPIE} for more details on such systematic errors.

 The parameters considered in all retrievals and the assumed priors are described in Table \ref{table:parameters}. Here, the priors are displayed as follows: $\mathcal{U}(x,y)$ represents a uniform (boxcar) prior with equal probability between a lower threshold $x$ and upper threshold $y$; $\mathcal{G}(\mu,\sigma)$ denotes a prior shaped like a Gaussian distribution with mean $\mu$ and standard deviation $\sigma$). The priors are generally consistent with our previous studies (see \pIIIaV), with differences in the assumed radius and mass priors. The radius prior is uniform between 0.5 and 2 $R_\oplus$, and the mass prior is a Gaussian of mean 1 and sigma 0.5 $M_\oplus$. These priors are wider than what we assumed in previous papers of the \life~series. While \life~will be able to gather narrower priors on the radius from the search campaign by measuring the emitted flux of the planet (see \pII~for details), the same assumptions could not be made in the case of reflected light measurements obtained with \hwo. To allow both concepts to be comparable fairly, we selected larger priors that are realistic estimates of what could be a product of prior measurements that precede a prior detailed characterization campaign.

As in \pIIIaV, in the retrieval we assume the presence of a filling gas whose abundance is $1-\Sigma (X_i)$ where $X_i$ is the abundance of the chemical species included as free parameters (see Table \ref{table:parameters}). The filling gas has a molecular weight of 28, assuming an \ce{N2}-dominated mixture which is realistic for terrestrial planets. The filling gas only contributes to the mean molecular weight of the atmosphere and it is not spectroscopically active. 

The retrieval also includes \ce{N2} as a free parameter to allow the inclusion of \ce{N2} Rayleigh scattering cross sections and \ce{N2}-related collision-induced absorption (see Table \ref{table:cia}). In principle, we could expect any gas (or mixture) of similar abundance and cross-section to be able to model the same Rayleigh feature in the reflection spectrum. The use of \ce{N2} in our retrievals is to be interpreted as a representation of a non-interacting bulk Rayleigh scatterer of an average molecular weight of 28. To ensure this distinction is clear, we label this parameter in our plots as \ce{X2}.

\subsection{Methods for the significance analysis}

Ultimately, when performing retrievals on either simulated or real data, we are interested in the significance of the detection of the main free parameters. In this work, we use established methodologies to quantify the significance of the detection of each atmospheric species with the single-instrument runs and the joint retrieval run.

\subsubsection{Bayes factor}

One well-known metric to quantify the robustness of a detection is the Bayes factor \citep[see, e.g.,][and references therein]{Parviainen2018}.  The Bayes factor analysis allows us to compare two models to determine which one best reproduces the data with the least amount of parameters. We can derive the Bayes factor by calculating the ratio of the Bayesian evidences $\mathcal{Z}(\mathbf{D}|\mathcal{M}_i)$ of two sets of models, given the data $\mathbf{D}$:

\begin{equation}
    \mathcal{K}=\frac{\mathcal{Z}(\mathbf{D}|\mathcal{M}_1)}{\mathcal{Z}(\mathbf{D}|\mathcal{M}_2)}\label{eq:bayes}
\end{equation}
By choosing as $\mathcal{M}_1$ ($\mathcal{M}_2$) a model that includes  (excludes) a specific species in the atmosphere and comparing the Bayesian evidences, we can determine if the presence of that atmospheric species as a free parameter in the model is justified by the data. The Jeffreys' scale \citep{Jeffreys:Theory_of_prob} allows us to then determine which model is preferred and with which confidence (see Table~\ref{table:jeffrey}).

\begin{table}
\caption{Jeffrey's scale \citep{Jeffreys:Theory_of_prob}.}
\label{table:jeffrey} 
\centering \setlength\extrarowheight{1pt} 
\begin{tabular}{cl}
\hline\hline 
$\log_{10}\left(\mathcal{K}\right)$ & Evidence Strength\\ 
\hline 
$<0$ &Support for $\mathcal{M}_2$\\
 $(0,0.5)$ &Very weak support for $\mathcal{M}_1$\\
 $[0.5,1)$ &Substantial support for $\mathcal{M}_1$\\
 $[1,2)$ &Strong support for $\mathcal{M}_1$\\
 $[2,\infty)$ &Decisive support for $\mathcal{M}_1$\\ 
\hline 
\end{tabular}
\tablefoot{Scale for the interpretation of the Bayes factor. Adapted from \pIIIaV.}
\end{table}

\subsubsection{``Sigma'' interpretation of the Bayes factor}

Following \citet{BennekeSeager2013} we are also able to calibrate between Bayesian and frequentist detections by relating the value of $\mathcal{K}$ to a level of significance. This conversion is possible provided the priors are either unimodal symmetric distributions or uniform, as it is in this study. In addition, the priors have to be the same for both models involved in the test. Then, the tests in the Bayesian and frequentist regimes become equivalent, since the priors cancel out \citep[see][]{2017arXiv170101467T}. However, the tests do not have the same meaning and interpretation due to the different frameworks. With the Bayesian interpretation, the posterior probability $P(\mathcal{M}_1\vert D)$ gives probabilistic support that $\mathcal{M}_1$ is more likely to be correct in comparison to $\mathcal{M}_2$, provided that the two models exhaust the space of possible models. On the other hand, the frequentist P-value gives the probability of observing a test statistic value that is larger or equal to the one we observe, if the null hypothesis $\mathcal{H}_0$ (here associated with $\mathcal{M}_2$) were to be correct. The P-value allows to reject $\mathcal{H}_0$ for a chosen type I error value. To ease up interpretation, the P-values are translated to a ``sigma'' value as frequently used in astronomy \citep[see, e.g.,][]{BennekeSeager2013}. This is to be interpreted as a complementary metric to the Bayes factor to determine the strength of the detection. In this case, only positive values are allowed, since no conclusive statement on the strength of the detection can be given when $\log_{10}(\mathcal{K})<0$ (i.e., when $\mathcal{M}_2$, the model that excludes that species, is preferred). 
In Table \ref{tab:Ksigma}, we show some relevant frequentist significance values and the corresponding Bayes factors and evidence strengths. We used these reference values to convert the $\mathcal{K}$ values into ``sigma''.
% \end{table}

\begin{table}[!tbh]

{\footnotesize
\caption{Translation table between some relevant frequentist significance values and their Bayesian interpretation. Adapted from \citet{BennekeSeager2013} and \citet{Jeffreys:Theory_of_prob}.}
\label{tab:Ksigma}
\begin{center}
    \setlength\extrarowheight{2pt} 

\begin{tabular}{@{\extracolsep{0.8pt}}llllll@{}}
\hline\hline
\multicolumn{4}{c}{  Bayesian}&\multicolumn{2}{c}{Frequentist}
\\
$\mathcal{K}$&$\log_{10}(\mathcal{K})$&$P(\mathcal{M}_1\vert D)$ &Ev. Strength &P-value&``sigma''\\
\cline{1-4} \cline{5-6}
2.5 & 0.40 & 0.714 & Weak & $0.05$  &       $2.0\sigma$ \\
2.9 & 0.46 &  0.744 & Weak&$0.04$  &       $2.1\sigma$\\
8.0 & 0.90 &  0.889 & Substantial& $0.01$ & $2.6\sigma$\\
12.0  &1.08 & 0.923 & Strong & $0.006$ &      $2.7\sigma$\\
21.0 & 1.32 & 0.955 & Strong& $0.003$   &    $3.0\sigma$\\
53.0&  1.72 & 0.981 & Strong &  $0.001$   &    $3.3\sigma$\\
150.0  &2.18 & 0.993 & Decisive & $0.0003$ &    $3.6\sigma$\\
43000.0 & 3.63 & 0.99998 &Decisive & $6\cdot10^{-7}$&  $5.0\sigma$\\
\hline
\end{tabular}\end{center}}
\end{table}

\subsubsection{Mean-squared error}

Since we are dealing with a simulated data scenario, we can exploit the knowledge of the true value to determine the precision and accuracy of the various retrievals. We use the mean-squared error, a metric that takes into account the variance of the posterior distribution and the distance of the mean of the posterior distribution from the true value.  The mean-squared error of an ``estimator'' parameter $\hat{\theta}$ compared to the true parameter  $\theta$ is defined as

\begin{equation}
    \mathrm{MSE}(\hat{\theta})=Bias(\theta,\hat{\theta})^2+Var(\hat{\theta})
\end{equation}
Where $Bias$ is the bias of the estimator, defined as the difference between the estimated value and the true value, and $Var$ is the variance of the estimator. In our scenario, this translates into:
\begin{equation}
    \mathrm{MSE}=(\mu-\tau)^2+\sigma^2
\label{eq:mse}\end{equation}

Where $\mu$ is the mean of the posterior distribution, $\tau$ is the true value, and $\sigma$ is the standard deviation of the posterior distribution. 

Since the priors assumed for various parameters span different ranges, we divided the posterior distribution of each parameter by the corresponding prior, so that the normalized posterior would become a value in the [0, 1] interval instead of the [x,y] interval.  For the mass, whose prior was Gaussian, we assumed the edges of the prior to be 0 and the 5$\sigma$ value from the mean of the Gaussian (3.5  $M_\oplus$). This allows us to represent all of the parameters with a single metric.

While this metric will not be available when dealing with real observations (since there will not be any true value to compare against), it can be useful at this stage of the development of the two future-generation concepts. By looking for the smallest $\mathrm{MSE}$, we can identify the scenario that provides the most precise and accurate estimates of relevant parameters. To better appreciate such small variations, we used the square root of the mean squared error for our analyses. A similar metric to evaluate the performance of different retrieval runs has been used in other recent works such as, for example, \citet{HayozCrocodile}.

\begin{table}
\caption{Details on the models run for each retrieval set and their color.} 
\label{tab:runcategories} 
    \centering
        \setlength\extrarowheight{1pt} 
    \begin{tabular}{p{0.5cm} p{3.2cm} p{4.3cm}}
    \hline\hline 
 \noindent Color & Range [$\mathrm{\mu}m$] & Description\\ 
\hline 
\noindent\parbox[c]{\hsize}{\crule[magenta]{0.4cm}{0.4cm}} & [0.2, 2.0] & Retrievals on \hwo~simulated spectra (\uvvisnir)\\
 \noindent\parbox[c]{\hsize}{\crule[blue]{0.4cm}{0.4cm}} & [4.0, 18.5]  & Retrievals on \life~simulated spectra (\mir) \\
 \noindent\parbox[c]{\hsize}{\crule[gold]{0.4cm}{0.4cm}} & [0.2, 2.0] $\cup$ [4.0, 18.5] & Joint retrievals on \hwo~and \life~(\uvvisnir+\mir)\\
 \hline
    \end{tabular}
\end{table}

\section{Results}\label{sec:results}

We performed retrievals on the sets described in Section \ref{sec:methods}. Each set in turn contains three retrievals performed on either \uvvisnir~data (\hwo-like), \mir~data (\life-like), or joint retrievals on \uvvisnir+\mir~data (\hwolife-like). These have been associated with the same colors throughout the manuscript (see Table \ref{tab:runcategories} for details). We describe our results for the idealized high-resolution low-noise scenario (used for validation) in Appendix \ref{sec:photonnoise}. Here, we focus on the simplified noise scenario in Section \ref{sec:constanterrorbars}, and the high-fidelity noise scenario in Section \ref{sec:scaledsnr}.

\subsection{Simplified noise scenario}\label{sec:constanterrorbars}

\begin{figure*}
     \centering
               \textbf{Simplified Noise}\par\medskip
\begin{subfigure}[b]{0.45\textwidth}
         \centering
         \includegraphics[width=\textwidth]{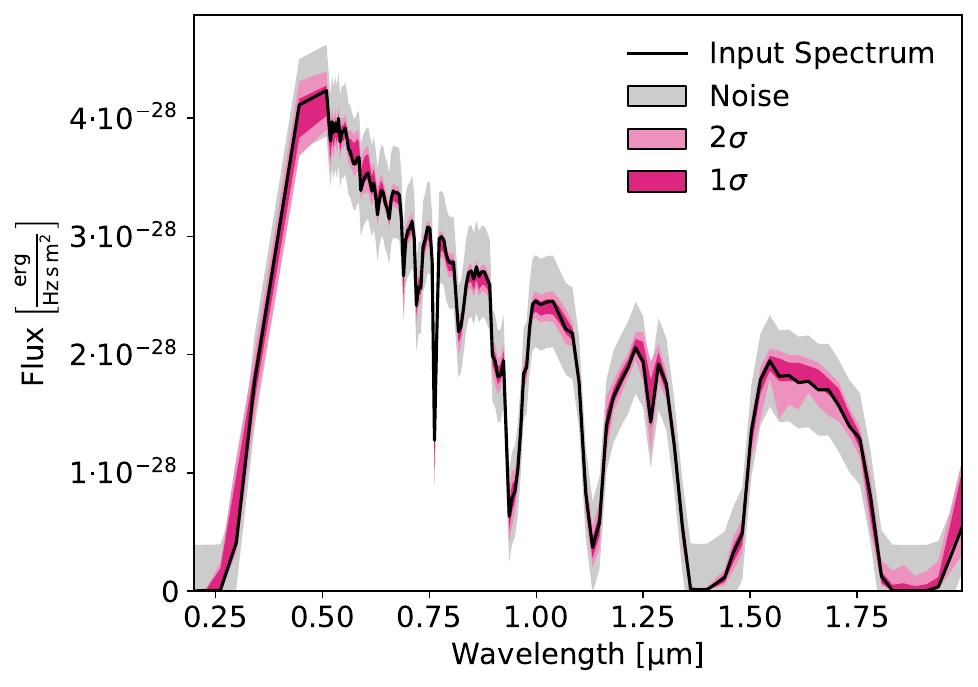}
     \end{subfigure}
          \begin{subfigure}[b]{0.45\textwidth}
         \centering
         \includegraphics[width=\textwidth]{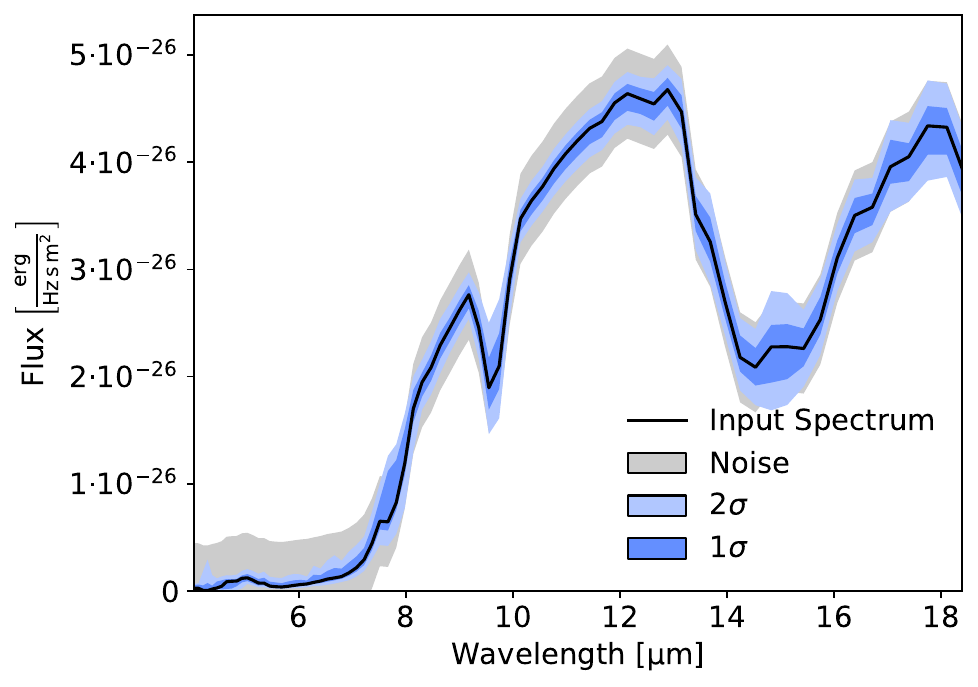}
     \end{subfigure}

     \begin{subfigure}[b]{0.92\textwidth}
         \centering
             \hspace{-0.4cm} \includegraphics[width=\textwidth]{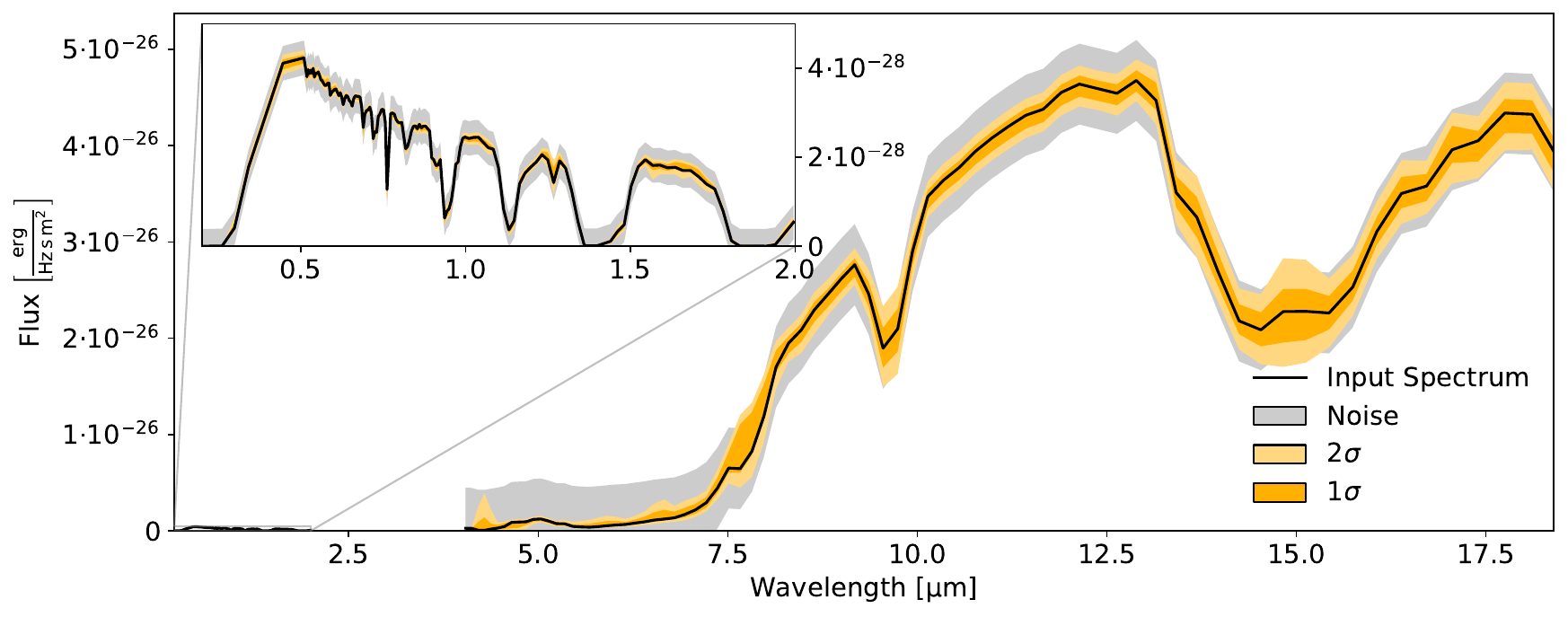}

     \end{subfigure}
          \begin{subfigure}[b]{0.5\textwidth}
         \centering
         \includegraphics[width=\textwidth]{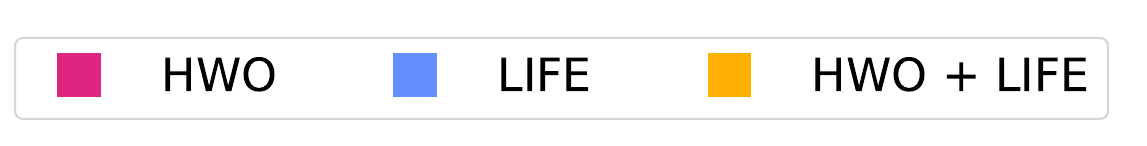}

     \end{subfigure}
        \caption{Retrieved spectra for the second retrieval set \emph{Left panel}: pure \hwo~retrieval (magenta);   \emph{Central panel}: pure \life~retrieval (blue); \emph{Right panel}: \hwolife~retrieval (yellow). In all panels, the 2-$\sigma$ and the 1-$\sigma$ intervals are shown in increasingly darker hues, as well as the input spectra (black lines) with error bars (gray-shaded areas) for comparison.}
        \label{fig:constanterrorbarsspectrum}
\end{figure*}

\begin{figure*}
     \centering
               \textbf{Simplified Noise}\par\medskip
     \begin{subfigure}[b]{0.33\textwidth}
         \centering
         \includegraphics[width=\textwidth]{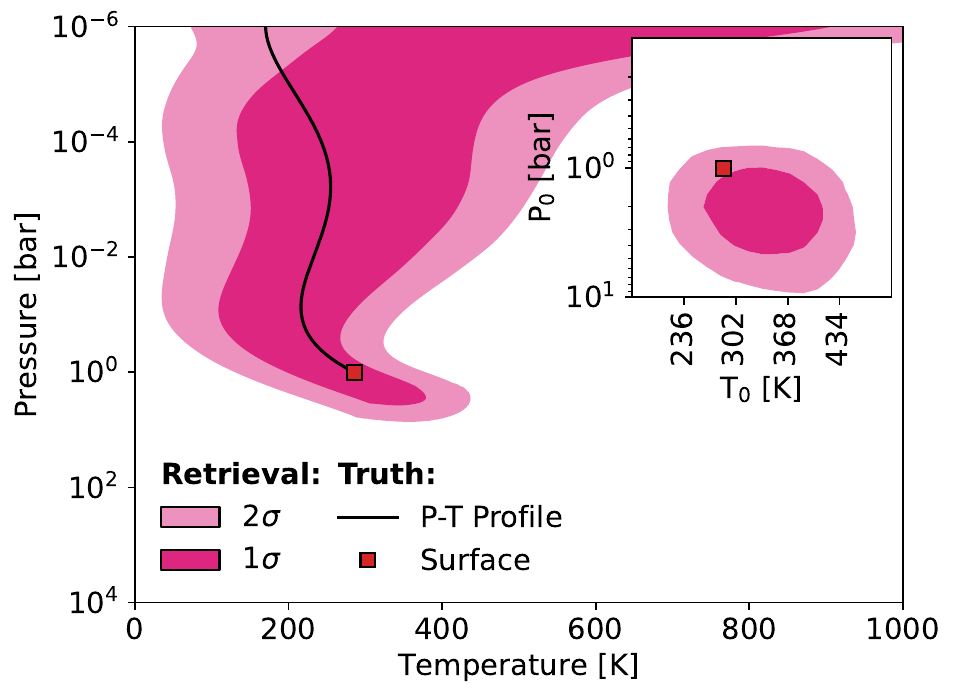}
     \end{subfigure}
     \hfill
     \begin{subfigure}[b]{0.33\textwidth}
         \centering
         \includegraphics[width=\textwidth]{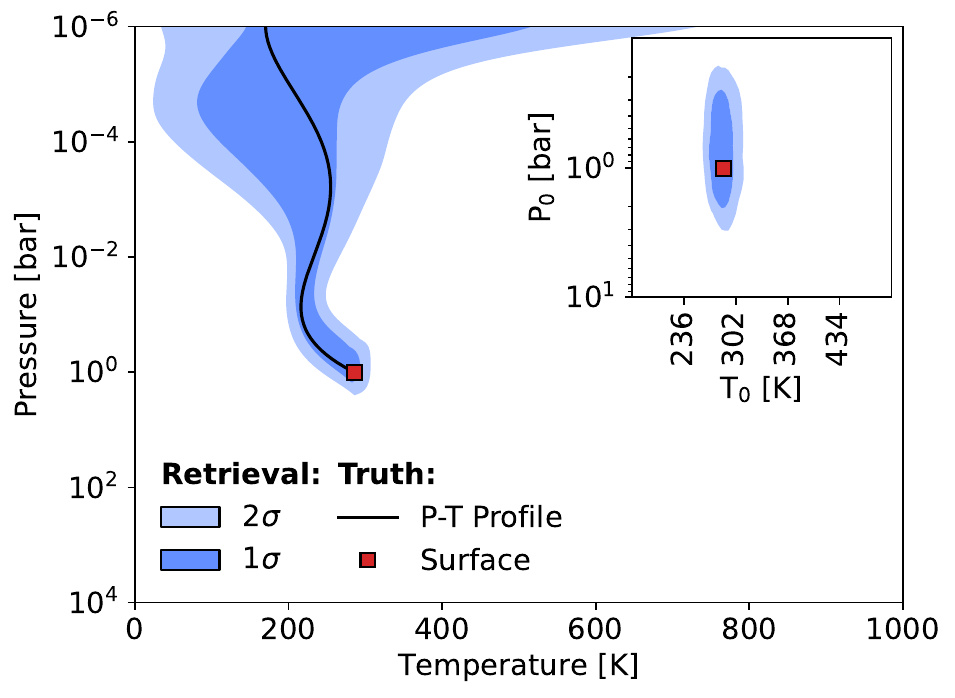}
     \end{subfigure}
\hfill
     \begin{subfigure}[b]{0.33\textwidth}
         \centering
         \includegraphics[width=\textwidth]{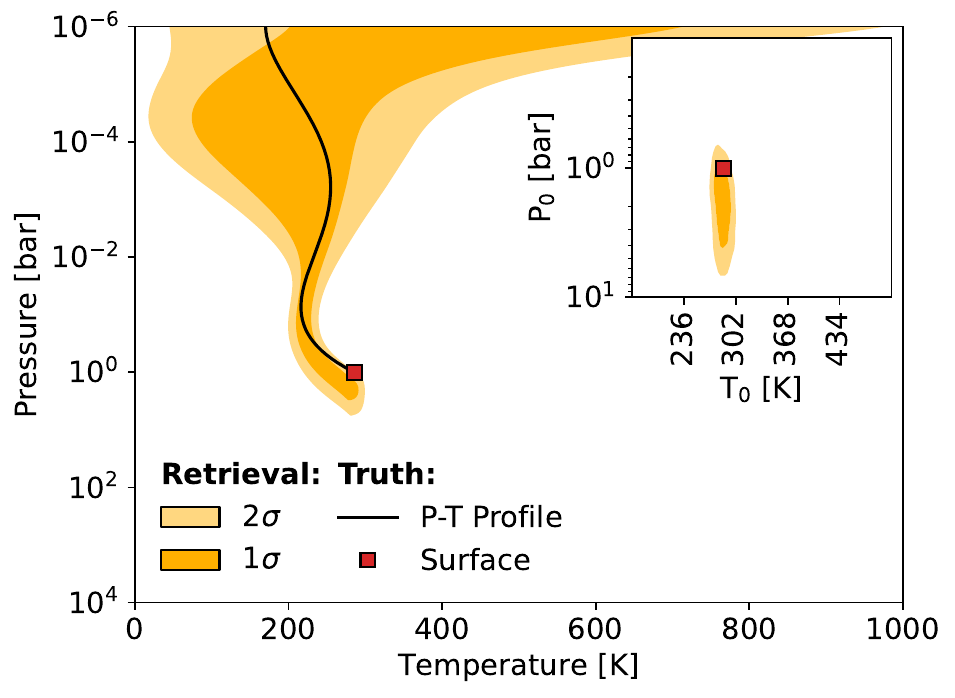}

     \end{subfigure}
          \begin{subfigure}[b]{0.5\textwidth}
         \centering
         \includegraphics[width=\textwidth]{Figures/legend.pdf}

     \end{subfigure}
        \caption{Retrieved pressure-temperature profiles for the second retrieval set (simplified noise): \emph{Left panel}: pure \hwo~retrieval (magenta);   \emph{Central panel}: pure \life~retrieval (blue); \emph{Right panel}: \hwolife~retrieval (yellow). In all panels, the 2-$\sigma$ and the 1-$\sigma$ intervals are shown in increasingly darker hues, as well as the input profiles (black lines) for comparison.  Inside each panel, the inset plot shows the 2D posterior space of the ground pressure and temperature. In all panels and the inset plots, the surface pressure and temperature point in the \pt~space is shown as a red square marker.}
        \label{fig:constanterrorbarspt}
\end{figure*}

\begin{figure}
\centering
 \textbf{Simplified Noise}\par\medskip
\includegraphics[width=\linewidth]{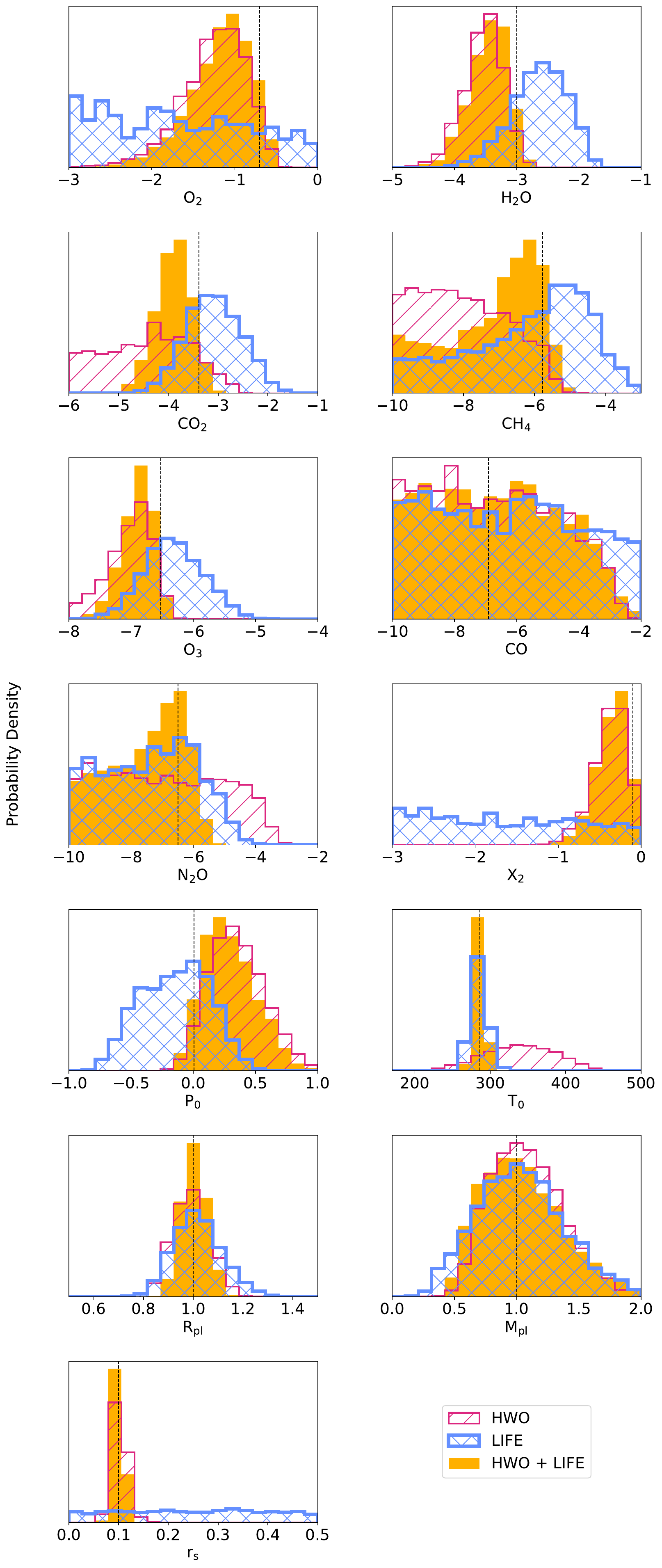}

\caption{Posterior density distributions from the second set of retrievals (simplified noise). The black lines indicate the expected values for every parameter. \hwo~posteriors are shown in magenta with diagonal hatching; \life~posteriors are shown in cyan with crossed hatching; \hwolife~posteriors are shown as fully colored gold histograms.}
\label{fig:constanterrorbarsposteriors}
\end{figure}

\begin{figure}
    \centering
     \textbf{Simplified Noise}\par\medskip
    \includegraphics[width=0.8\linewidth]{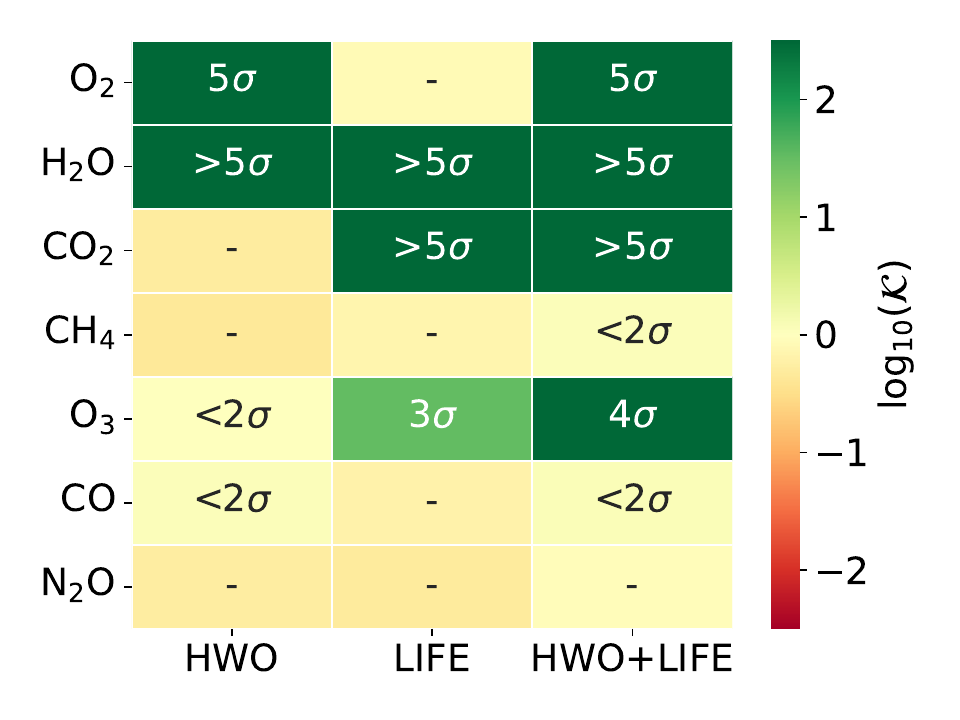}
    \caption{Bayes factor and confidence levels values for each retrieved spectroscopically active species in the second set of retrievals (simplified noise). The heatmap is color-coded according to the Bayes factor values $\log_{10}(\mathcal{K})$ and the confidence levels (in ``sigma'' values) are labeled within each cell whenever $\log_{10}(\mathcal{K})$ is positive. The Bayes factor values were obtained by performing a retrieval including and excluding the molecule of interest and then using Eq.~\ref{eq:bayes}. These were then converted into $\sigma$ through Table~\ref{tab:Ksigma}. The Bayes factor values can be found in Table \ref{apptab:sigmaconstanterrorbar}.}
    \label{fig:sigmaconstanterrorbar}
\end{figure}

\begin{figure}[h]
     \centering
               \textbf{Simplified Noise}\par\medskip
     \begin{subfigure}[b]{\linewidth}
         \centering
         \includegraphics[width=0.9\textwidth]{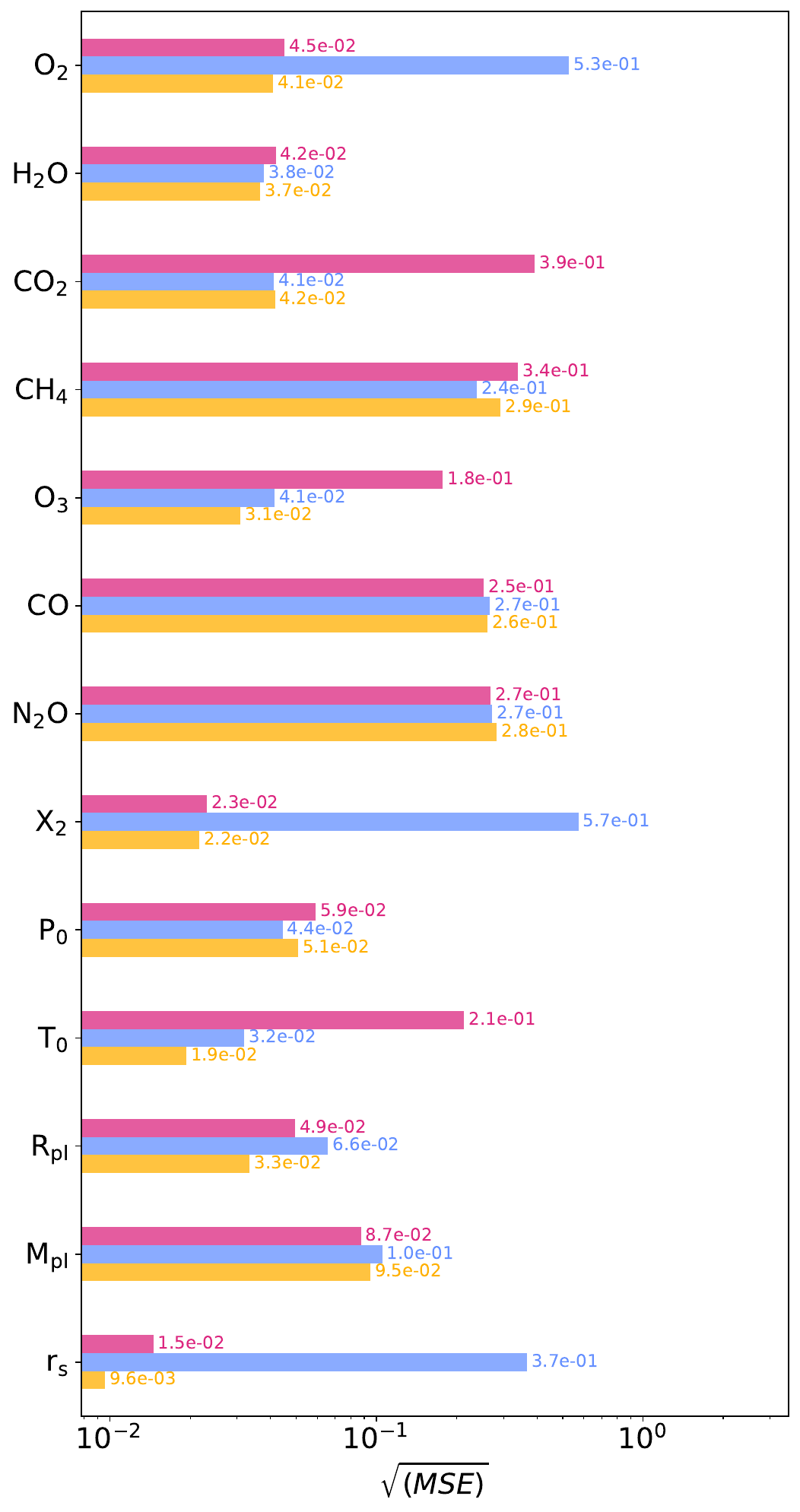}
     \end{subfigure}

          \begin{subfigure}[b]{0.5\textwidth}
         \centering
         \includegraphics[width=\textwidth]{Figures/legend.pdf}

     \end{subfigure}
        \caption{Square root of the mean squared error (see Equation \ref{eq:mse}) for relevant parameters in the second set of retrievals (simplified noise).}
        \label{fig:constanterrorbarsmse}
\end{figure}

In this set of retrievals, we assume the current baseline resolutions for \life~and \hwo. These are: \Rv{50} from \mic{4-18.5} for \life; \Rv{7} (\mic{0.2-0.515}), \Rv{140} (\mic{0.515-1})  and \Rv{70} (\mic{1.01-2}) for \hwo. 
For this set of retrievals, we keep a simplified noise instance (for more details see Section \ref{sec:observedspectra}). 
In Figures \ref{fig:constanterrorbarsspectrum}, \ref{fig:constanterrorbarspt},  and \ref{fig:constanterrorbarsposteriors} we show a comparison of the retrieved spectra, the \pt~profiles, and the posteriors for the three models in this set. 
The Bayes factors and confidence levels in $\sigma$ values are shown in Figure \ref{fig:sigmaconstanterrorbar} and Table \ref{apptab:sigmaconstanterrorbar}. The values of the square root of the $\mathrm{MSE}$ are shown as bar plots in Figure~\ref{fig:constanterrorbarsmse}. The corner plot and a table containing the retrieved estimates with their 1-$\sigma$ uncertainty for each parameter can be found in Figure \ref{appfig:constanterrorbarscorner}.

As shown in Figure \ref{fig:constanterrorbarsspectrum}, the retrieved spectrum is always within the noise uncertainty (gray shaded area) for all models. For the single-instrument retrievals, the retrieved spectra show larger uncertainties at short wavelengths where the noise is higher, as well as in some prominent lines. Such uncertainties on the retrieved spectrum are smaller in the \hwolife~retrieval, as we would expect from a retrieval performed on a larger amount of data which exploits a larger information content compared to the retrievals on a single wavelength range. 

Regarding the characterization of the thermal structure of the atmosphere, the retrieval on \life~data infers a more precise and accurate characterization of the atmosphere, though with increasing uncertainties towards lower pressures. This is related to the physics of the problem: as the majority of the radiation in the thermal emission spectrum comes from the deeper layers of the atmosphere and the surface, it is much easier to constrain the deeper layers of the atmosphere rather than the upper layers.
The retrieved value of the ground pressure is around $\log_{10}(\Ps)=-0.13$ which corresponds to about 0.74 bar, with an uncertainty of $\approx$0.3 dex. The true surface pressure value is within the 1-$\sigma$ uncertainty. The surface temperature is estimated as $\Ts=284\pm 10$ K, very close to the true value of 286 K.

The \hwo~run does not obtain the same accuracy on the thermal profile, with much wider uncertainties across all pressures. The ground pressure is overestimated as $\log_{10}(\Ps)=0.33\pm 0.25$, which corresponds to $\approx$ 2.1 bar, a factor two larger than the expected value. 
The surface temperature is also much more loosely constrained at $\Ts=335\pm 55$ K, about 50 K hotter than the true value.

For the joint \hwolife~retrieval, the surface pressure is also slightly overestimated (retrieved $\log_{10}(\Ps)=0.27\pm0.2$, which corresponds to around 1.8 bar), though less than the \hwo-only run.

The ground temperature retrieval obtained with a joint retrieval is more precise and accurate than any of the single-instrument retrievals, with a retrieved ground temperature of $\Ts=286\pm6$ K. The \pt~profile is also retrieved correctly, with similar uncertainties as the \life~retrieved profile (see Figure \ref{fig:constanterrorbarspt}) at lower pressures, but even smaller uncertainties in the deeper layers of the atmosphere. 

When it comes to the characterization of the mass and radius, there is no substantial difference between the various runs (see Figures \ref{fig:constanterrorbarsposteriors} and \ref{appfig:constanterrorbarscorner}). The radius is accurately retrieved by all models with an uncertainty of 0.1 $R_\oplus$ or less. Specifically, the retrieval of \life~and \hwo~are comparable with each other, while the \hwolife~one gets a more accurate constraint on the radius up to 0.05 $R_\oplus$ uncertainty. The mass is consistently retrieved in all three retrievals, but its estimate is not significantly more precise than the input prior. 

Regarding the reflectance, the \hwo~and \hwolife~retrievals correctly retrieve the value with comparable uncertainty, while \life~alone does not manage to constrain this parameter. This shows that only the reflected-light portion of the spectrum contains enough information to correctly retrieve the reflectance of the planet, which is to be expected and is confirmed by the idealized high-resolution low-noise scenario (see Appendix \ref{sec:photonnoise}).

By observing the posteriors in Figure \ref{fig:constanterrorbarsposteriors} we note that, generally, the posteriors of the joint retrievals are significantly smaller than the posteriors of the single models, especially when the \life~and \hwo~retrievals are not consistently retrieving a parameter. The \hwolife~posteriors place themselves at the intersection of the posteriors of the single-instrument runs; this translates into smaller uncertainties on the estimates of most parameters and a greater decoupling of all correlated parameters. In other words, compared to single-instrument retrievals, a joint retrieval reduces the subset of possible parameters that can reproduce the data, particularly so when two or more parameters are correlated (e.g., the pressure and the main absorbing species of the atmosphere). This means that the joint retrieval correctly leverages the total amount of information available in the two wavelength ranges, as one would expect from the retrieval of complementary data.

Figures \ref{fig:constanterrorbarsposteriors}, \ref{fig:sigmaconstanterrorbar}, and \ref{fig:constanterrorbarsmse} quantify what we observed so far. 
The retrieval of the atmospheric species with \life~is consistent with the previous papers in the series: \life~can decisively constrain the abundance of \ce{H2O} and \ce{CO2} (with more than 5$\sigma$ confidence level), as well as strongly constrain \ce{O3} ($3\sigma$ confidence). However, it is not sensitive to the bulk scatterer \ce{X2}, \ce{O2}, and \ce{CO}, and can only retrieve upper limits on \ce{N2O} and \ce{CH4}.
On the other hand, the \hwo~run poses strong constraints on \ce{O2} and \ce{H2O} ($5\sigma$), while \ce{O3} is weakly constrained (at a $<2\sigma$ confidence level). The other molecules (\ce{CH4}, \ce{CO}, \ce{CO2}, \ce{N2O}) are unconstrained or constrained with very broad upper limits. Through \hwo~retrievals, it is also possible to estimate the abundance of the bulk absorber \ce{X2} (see Section \ref{sec:singleinstruments}).
The joint retrieval can retrieve with higher precision the species that were already retrieved in both the single-instrument retrievals. For all the molecules we retrieved, the \hwolife~retrieval obtains equal or higher confidence levels (Figure \ref{fig:sigmaconstanterrorbar}) compared to the single-instrument cases. Notably, it is possible to have a decisive detection of \ce{O3} compared to a weak and strong detection of the \hwo~and \life~retrievals respectively. This result is however dependent on the noise simulation, as will be seen in the third set of retrievals and further discussed in Section \ref{sec:noise}.

The mean-squared-error metric is also valuable to quantify the improvement of the joint retrieval compared to the single-instrument scenarios (Figure \ref{fig:constanterrorbarsmse}): for all the parameters considered in the retrieval, the estimate provided by the multi-instrument retrieval is comparable or significantly better in precision and accuracy compared to the single-instrument estimates (i.e., the value of $\sqrt{\mathrm{MSE}}$ of the \hwolife~retrieval for each parameter is the same or lower than the one from the single-instrument retrievals).

\subsection{Higher-fidelity PSG/\lifesim~simulated noise scenario} \label{sec:scaledsnr}

\begin{figure*}[!htbp]
     \centering
    \textbf{PSG/\lifesim~Noise}\par\medskip
\begin{subfigure}[b]{0.45\textwidth}
         \centering
         \includegraphics[width=\textwidth]{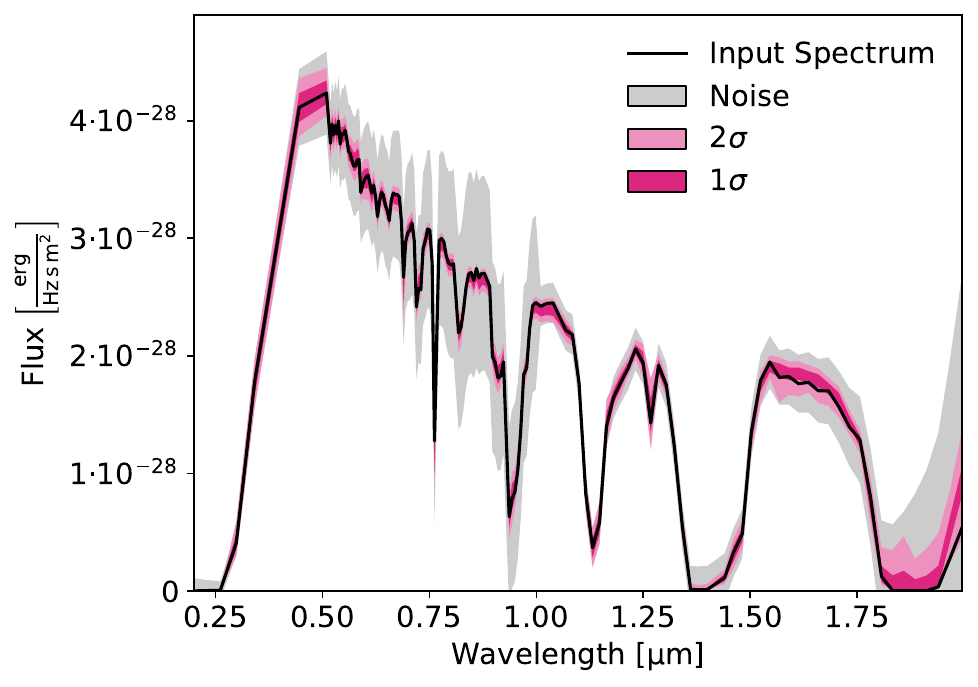}
     \end{subfigure}
          \begin{subfigure}[b]{0.45\textwidth}
         \centering
         \includegraphics[width=\textwidth]{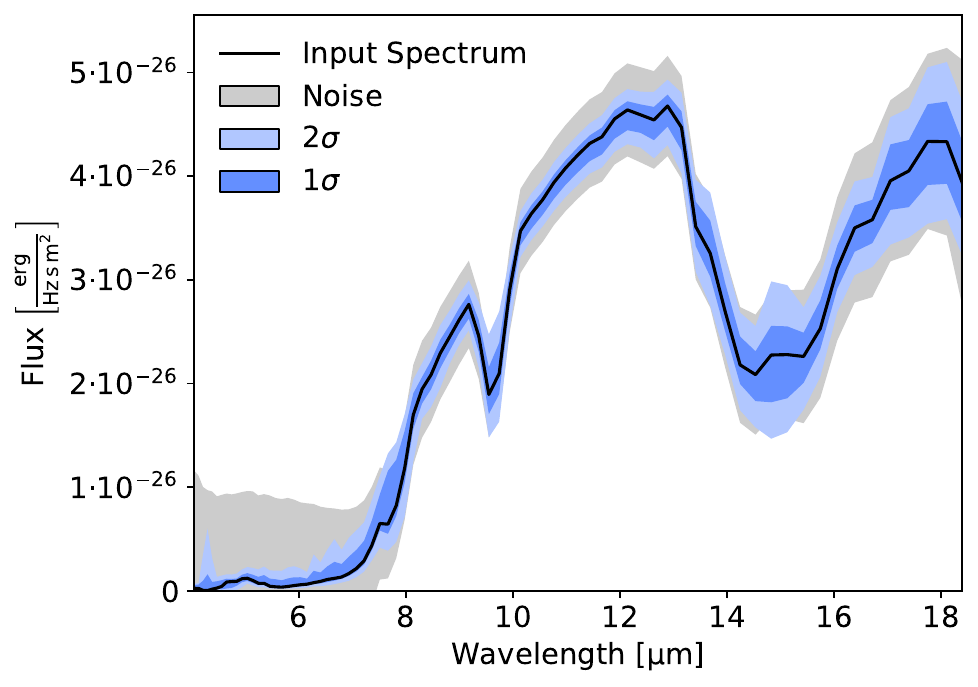}
     \end{subfigure}

     \begin{subfigure}[b]{0.92\textwidth}
         \centering
             \hspace{-0.4cm} \includegraphics[width=\textwidth]{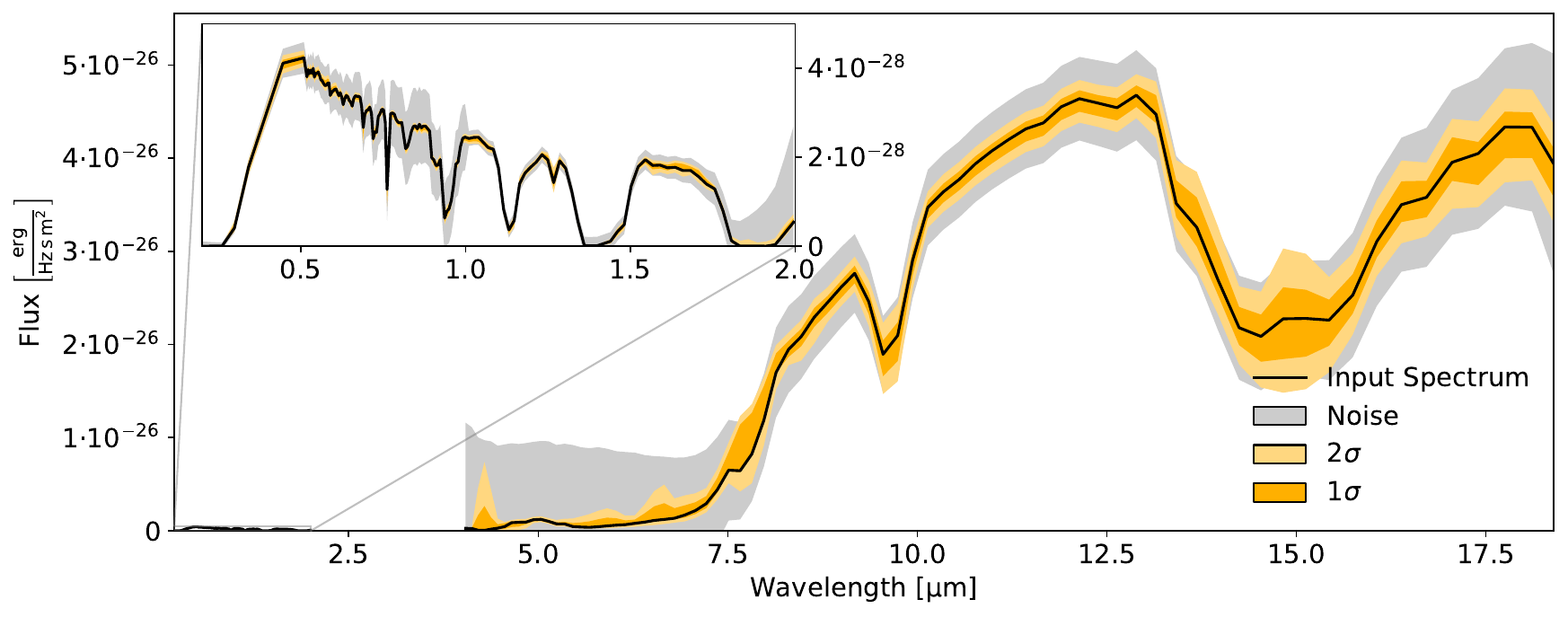}

     \end{subfigure}
           \begin{subfigure}[b]{0.5\textwidth}
     \centering
     \includegraphics[width=\textwidth]{Figures/legend.pdf}

 \end{subfigure}
        \caption{Retrieved spectra for the third retrieval set (PSG/\lifesim~noise): \emph{Left panel}: pure \hwo~retrieval (magenta);   \emph{Central panel}: pure \life~retrieval (blue); \emph{Right panel}: \hwolife~retrieval (yellow). In all panels, the 2-$\sigma$ and the 1-$\sigma$ intervals are shown in increasingly darker hues, as well as the input spectra (black lines) with error bars (gray-shaded areas) for comparison.}
        \label{fig:scaledsnrspectrum}
\end{figure*}

\begin{figure*}[!htbp]
     \centering
    \textbf{PSG/\lifesim~Noise}\par\medskip
               
\begin{subfigure}[b]{0.33\textwidth}
     \centering
     \includegraphics[width=\textwidth]{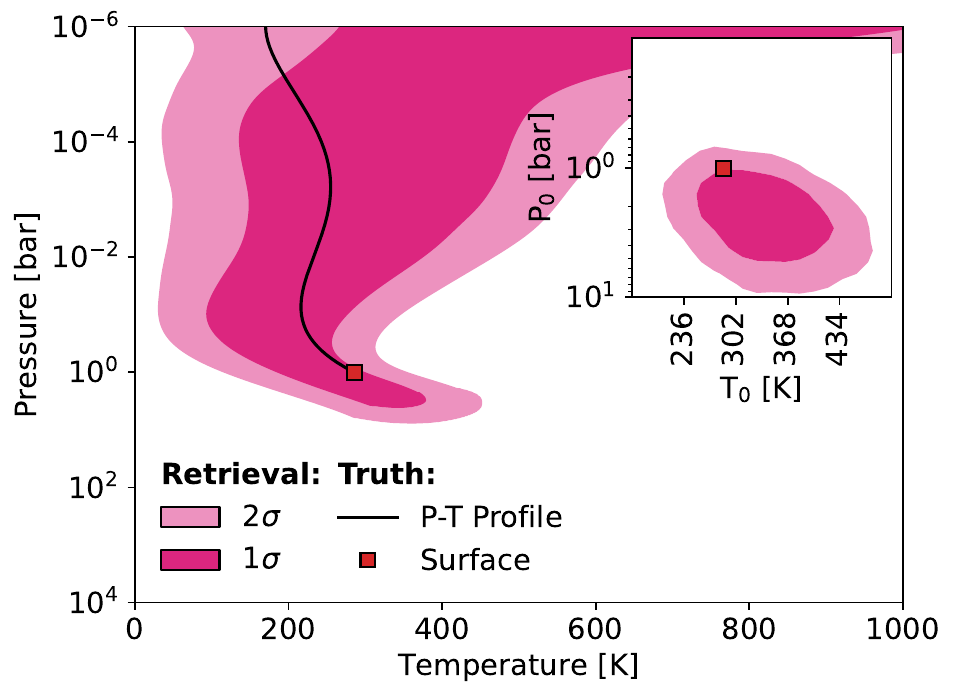}
 \end{subfigure}
 \hfill
 \begin{subfigure}[b]{0.33\textwidth}
     \centering
     \includegraphics[width=\textwidth]{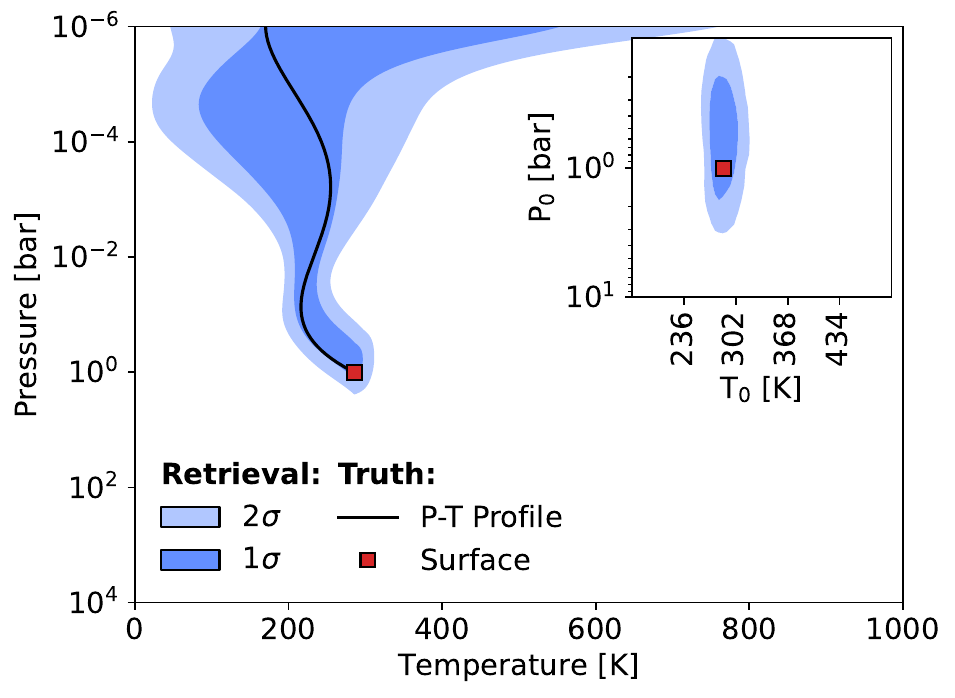}
 \end{subfigure}
\hfill
 \begin{subfigure}[b]{0.33\textwidth}
     \centering
     \includegraphics[width=\textwidth]{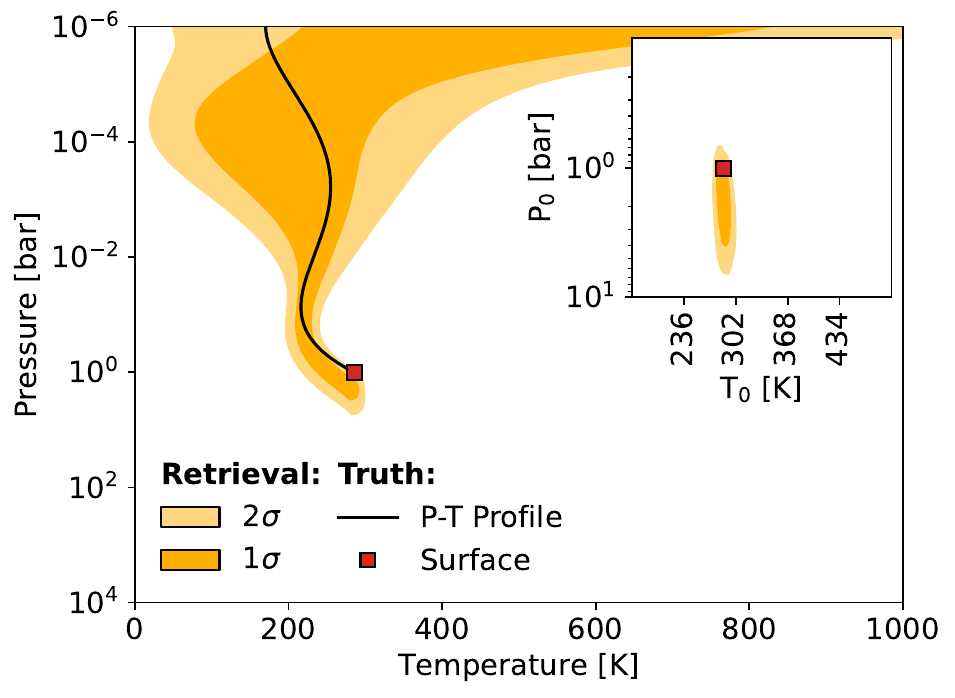}

 \end{subfigure}
      \begin{subfigure}[b]{0.5\textwidth}
     \centering
     \includegraphics[width=\textwidth]{Figures/legend.pdf}

 \end{subfigure}
    \caption{Retrieved pressure-temperature profiles for the third retrieval set (PSG/\lifesim~noise): \emph{Left panel}: pure \hwo~retrieval (magenta);   \emph{Central panel}: pure \life~retrieval (blue); \emph{Right panel}: \hwolife~retrieval (yellow). In all panels, the 2-$\sigma$ and the 1-$\sigma$ intervals are shown in increasingly darker hues, as well as the input profiles (black lines) for comparison.  Inside each panel, the inset plot shows the 2D posterior space of the ground pressure and temperature. In all panels and the inset plots, the surface pressure and temperature point in the \pt~space is shown as a red square marker.}
        \label{fig:scaledsnrpt}
\end{figure*}

\begin{figure}
\centering
 \textbf{PSG/\lifesim~Noise}\par\medskip
 \includegraphics[width=\linewidth]{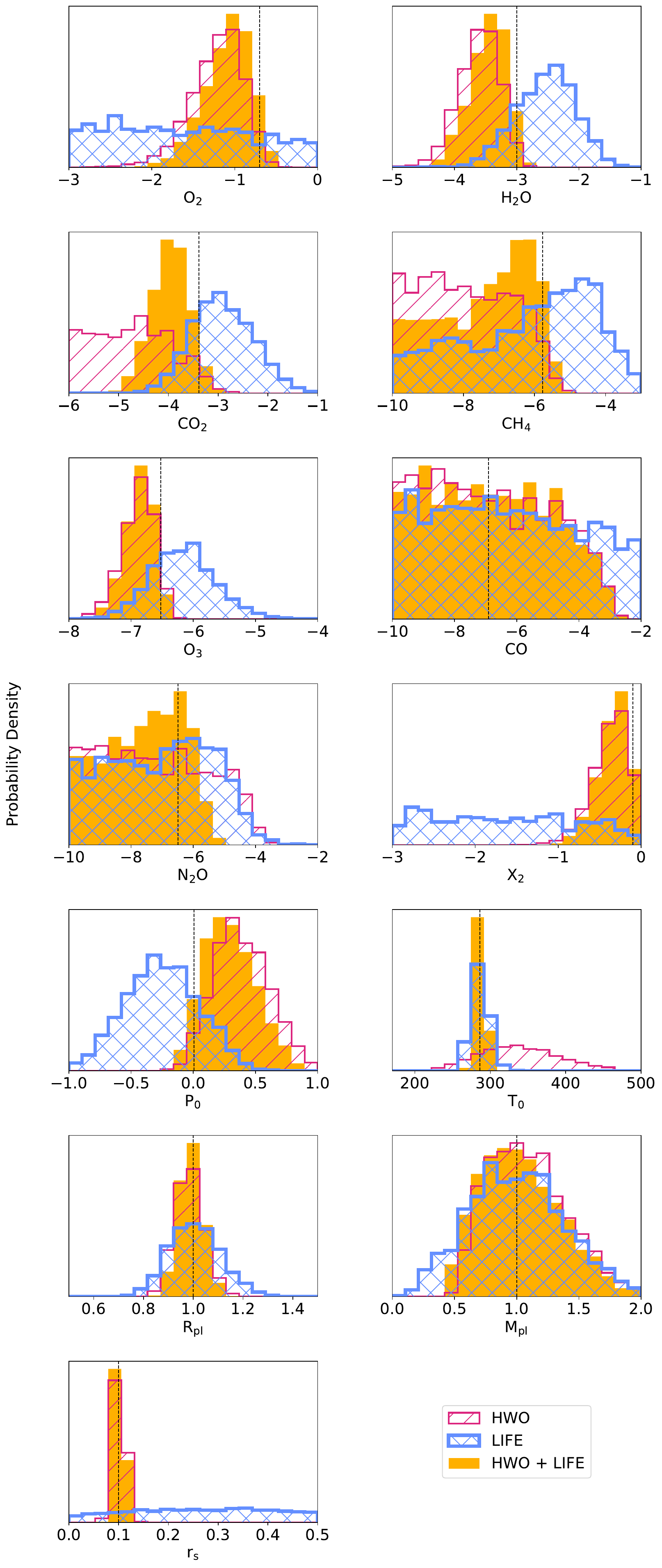}

\caption{Posterior density distributions from the third set of retrievals (PSG/\lifesim~noise). The black lines indicate the expected values for every parameter. \hwo~posteriors are shown in magenta with diagonal hatching; \life~posteriors are shown in cyan with crossed hatching; \hwolife~posteriors are shown as fully colored gold histograms.}
\label{fig:scaledsnrsposteriors}
\end{figure}

\begin{figure}
    \centering
    \textbf{PSG/\lifesim~Noise}\par\medskip
    \includegraphics[width=0.8\linewidth]{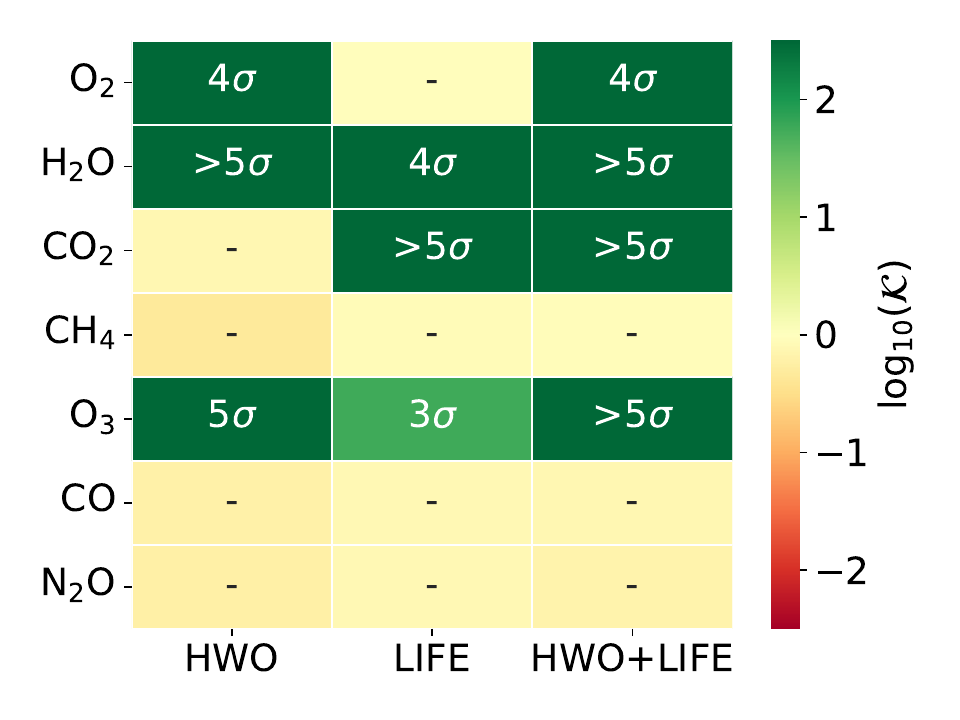}
    \caption{Bayes factor and confidence levels values for each retrieved spectroscopically active species in the second set of retrievals (PSG/\lifesim~noise). The heatmap is color-coded according to the Bayes factor values $\log_{10}(\mathcal{K})$ and the confidence levels (in ``sigma'' values) are labeled within each cell whenever $\log_{10}(\mathcal{K})$ is positive. The Bayes factor values were obtained by performing a retrieval including and excluding the molecule of interest and then using Eq.~\ref{eq:bayes}. These were then converted into $\sigma$ through Table~\ref{tab:Ksigma}. The Bayes factor values can be found in Table \ref{apptab:sigmascaledsnr}.}
    \label{fig:sigmascaledsnr}
\end{figure}

\begin{figure}
     \centering
               \textbf{PSG/\lifesim~Noise}\par\medskip
     \begin{subfigure}[b]{\linewidth}
         \centering
         \includegraphics[width=0.9\textwidth]{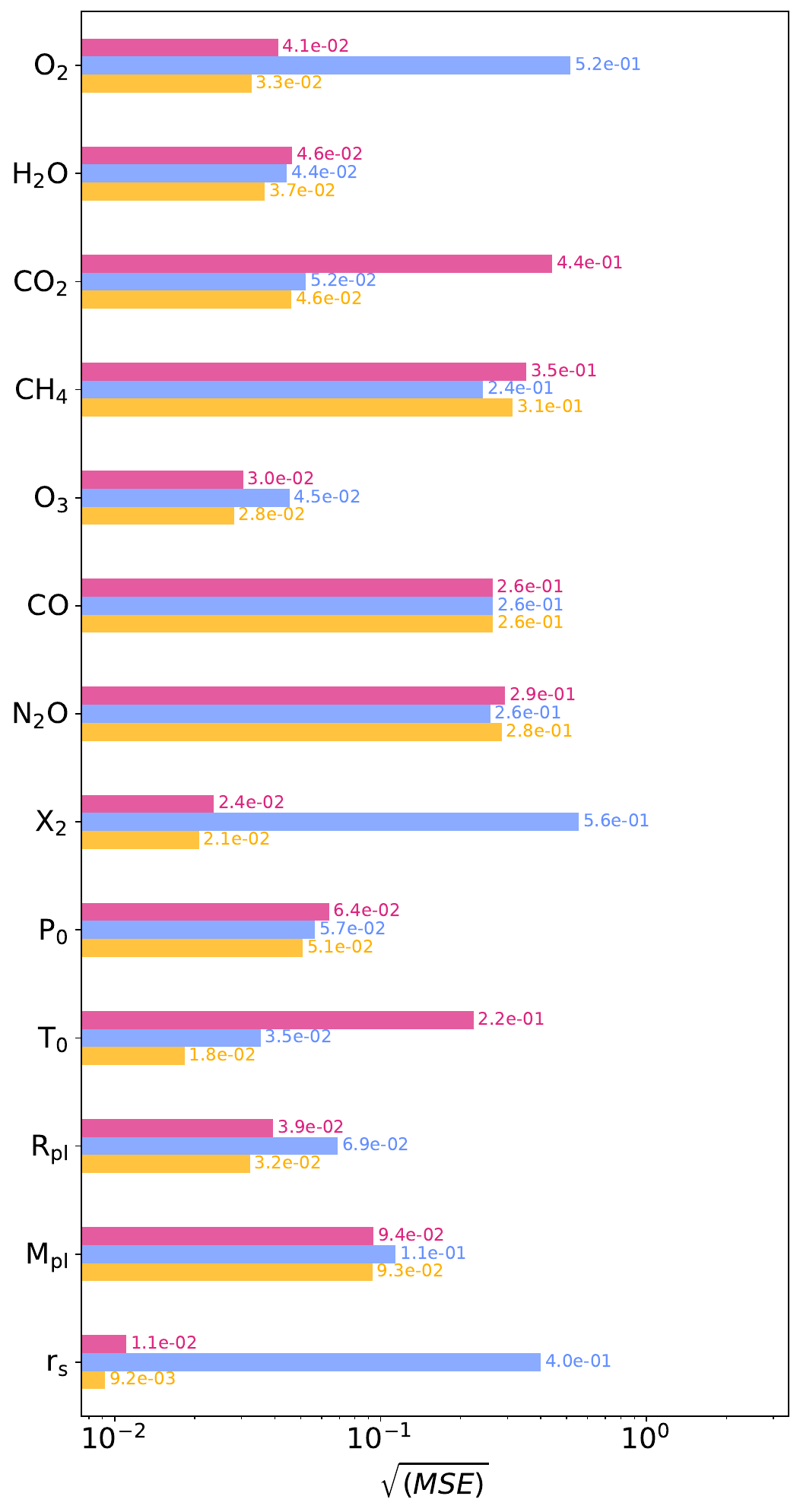}
     \end{subfigure}

          \begin{subfigure}[b]{0.5\textwidth}
         \centering
         \includegraphics[width=\textwidth]{Figures/legend.pdf}

     \end{subfigure}
        \caption{Square root of the mean squared error (see Equation \ref{eq:mse}) for relevant parameters in the third set of retrievals (PSG/\lifesim~noise).   }
        \label{fig:scaledsnrmse}
\end{figure}

In this set of retrievals, we used the same ``baseline'' resolutions as the previous set but assumed a more complex noise instance, which was generated by the currently available noise simulators for the \life~and the \hwo~concepts (as described in Section \ref{sec:observedspectra}). 
The results of this set of retrievals are shown in Figures \ref{fig:scaledsnrspectrum}, \ref{fig:scaledsnrpt}, and \ref{fig:scaledsnrsposteriors}.  The Bayes factor and confidence level values of the atmospheric species are shown in Figure \ref{fig:sigmascaledsnr}. The numeric values of the Bayes factor are reported in Table \ref{apptab:sigmascaledsnr}. The square root of the mean squared error for the retrieved parameters is shown in Figure \ref{fig:scaledsnrmse}. The corner plot and the retrieved estimates with their 1-$\sigma$ uncertainty for each parameter can be found in Figure \ref{appfig:scaledsnrcorner}.

The \pt~profile is accurately retrieved with LIFE. On the other hand, the \hwo~run overestimates the ground pressure and temperature and constrains the atmospheric profile more loosely. The joint retrieval overestimates the ground pressure, but still retrieves a more accurate and precise estimate of the ground temperature compared to the single-instrument runs ($\Ts=287\pm 6$~K for the \hwolife~run compared to $\Ts=336\pm60$~K for the \hwo~run and $\Ts=285\pm11$~K for the \life~run, compared to the true value of 286 K).

Regarding the atmospheric characterization (see Figures \ref{fig:scaledsnrsposteriors}, \ref{fig:sigmascaledsnr}, and \ref{fig:scaledsnrmse}) we observe some differences compared to the previous set of models. The retrieval on \hwo~data estimates with strong or decisive confidence ($\geq3\sigma$ confidence level) the abundances of \ce{O2}, \ce{H2O}, and \ce{O3}. The confidence on the \ce{O3} detection is a definite improvement compared to the previous set, though at the expense of a lack of confidence in the retrieval of an upper limit of \ce{CO}. 

The performance of the \life~retrieval does not significantly change in this scenario.
 Also for this set of retrievals, the joint retrieval of \uvvisnir~+ \mir~data produces results at least as confident as the single-instrument runs, estimating all free parameters with higher accuracy and precision compared to \hwo- and \life-only retrievals (see Figure \ref{fig:scaledsnrmse}). 
 However, the joint retrieval is not able to confidently estimate \ce{CH4}, \ce{CO}, or \ce{N2O}.

\section{Discussion}\label{sec:discussion}

In this section we discuss the results, focusing on the single-instrument observations (Section \ref{sec:singleinstruments}), the joint observations (Section \ref{sec:complementarity}), and the impact of noise (Section \ref{sec:noise}). We discuss the limitations of our work in Section \ref{sec:limitations}. To benchmark our results with the literature whenever possible, we compare the single-instrument runs we performed in this work with other \hwo- and \life-centered studies in Appendix \ref{app:comparison}.

\subsection{Strengths of single-instrument observations}\label{sec:singleinstruments}

In all the retrieval sets performed in this study, we can directly compare the potential of the two mission concepts in characterizing a Modern Earth-twin exoplanet. Most of the differences lie in the intrinsic planetary spectral features that are observable in different wavelength bands. Molecular oxygen has sharp features in the visible reflected-light spectrum, while ozone is prominent both in the UV (<\mic{0.3}) and in the \mir~(around \mic{9.6}). The \ce{CO2} band at \mic{15} is also a readily detectable feature, while \ce{H2O} absorbs across most wavelengths. Because of these differences, we can retrieve with higher confidence specific molecules in the two wavelength ranges (see Figures \ref{fig:constanterrorbarsposteriors}, \ref{fig:scaledsnrsposteriors}, and \ref{fig:validationposteriors}, as well as Tables \ref{apptab:sigmaconstanterrorbar}, \ref{apptab:sigmascaledsnr}). In all baseline scenarios (simplified noise and high-fidelity noise), from the \hwo~retrievals we performed we can estimate \ce{O2}, \ce{H2O}, and \ce{O3}; on the other hand, \life~retrievals allow us to estimate \ce{CO2}, \ce{H2O} and \ce{O3}. In both wavelength ranges, we also have upper limits on \ce{CH4} and \ce{N2O} (see Figures \ref{fig:constanterrorbarsposteriors} and \ref{fig:scaledsnrsposteriors}), though with very weak or no confidence for the baseline values. This is in agreement with what was found in the literature and our previous studies (see Appendix \ref{app:comparison}). Importantly, even in the idealized scenario (photon noise only, \Rv{1000} and \SNv{50}, see Appendix \ref{sec:photonnoise}), the single-instrument runs would still not be able to constrain \ce{CO} and \ce{N2O} for the \hwo~case, or \ce{O2} for the \life~case (see Figure \ref{appfig:validationcorner}).

Observations in the \uvvisnir~are especially sensitive to the scattering processes (both Rayleigh and surface scattering), while \mir~measurements would directly retrieve an estimate of the thermal structure of the atmosphere. These are results that can be observed clearly in all the runs (see, e.g., Figures  \ref{fig:constanterrorbarsposteriors}, \ref{fig:scaledsnrsposteriors}, and \ref{fig:validationposteriors}). 
Specifically, in the \hwo~retrievals the  Rayleigh slope is well reproduced by the forward model (see, e.g., Figures \ref{fig:constanterrorbarsspectrum}, \ref{fig:scaledsnrspectrum}, and \ref{fig:validationspectrum}). The Rayleigh slope in the \uvvisnir~spectrum is mainly impacted by the most abundant molecules, which could be spectrally active (e.g., \ce{O2}) or inactive (e.g., the bulk scatterer \ce{X2}, which would correspond to \ce{N2} in an Earth-twin scenario). While the abundance of spectrally active species can be constrained through absorption lines in the spectrum, \ce{X2} can only be constrained through the modeling of the Rayleigh slope. With \hwo-like observations, we would be able to infer the presence of a bulk scatterer \ce{X2}, though further studies would be needed to properly assess the nature of the absorber itself \citep[see][]{Hall2023}.

A correct retrieval of \ce{X2} and its contribution to scattering at short wavelengths in turn breaks the degeneracy between radius and reflectance: at short wavelengths, the planetary spectrum would become opaque because of Rayleigh scattering, so most of the incident flux never reaches the surface \citep[see, e.g.,][]{Gaudi2020} -- this allows these parameters to be retrieved with greater accuracy. On the other hand, retrievals of data at \uvvisnir~wavelengths cannot constrain the pressure-temperature profile to more than tens of degrees of uncertainty, even in the best-case scenario (see Figure \ref{fig:validationpt}). This result informs us that there is simply not enough information in the reflected-light spectrum to accurately pinpoint the thermal profile of the atmosphere. At the baseline resolutions, \hwo-like retrievals have the potential of also overestimating the ground pressure (see Figures \ref{fig:constanterrorbarspt} and \ref{fig:scaledsnrpt}). This biased result is coupled with a lower estimate of the abundance of some molecules in the \hwo-only runs (e.g., \ce{O2}, \ce{H2O}, \ce{O3}), as a result of the well-known pressure-abundance degeneracy, observed in many previous studies (see Appendix~\ref{app:comparison}). 

From the \mir~point of view, strong absorption lines of \ce{CO2}, \ce{H2O} and \ce{O3} allow for precisely fitting the \pt~profile in the denser layer of the atmosphere (see, e.g., Figures \ref{fig:constanterrorbarspt}, \ref{fig:scaledsnrpt}, and \ref{fig:validationpt}). \life~retrievals fail to constrain the surface reflectance $r_s$. This is expected since the stellar and thermal radiation that is scattered by the surface at \mir~wavelengths is so little to be below the noise level. The only exception is the high-resolution case: the impact of the scattered light on the spectrum is greater than the noise level for surface reflectances higher than 0.3-0.4, which allows the retrieval framework to provide an upper limit on this parameter for \mir-only data. Still, other estimates of the albedo of the planet are accessible through \life-like retrievals. For example, from an accurate estimate of the radius, which \life~would gather from the search campaign, it would be possible to get an estimate of the equilibrium temperature of the planet and, as a consequence, the Bond albedo of the planet \citep[see, e.g.,][i.e., \life~Paper IX, for details]{Konrad2023}.

When it comes to the interpretation of these results, single-instrument retrievals would allow us to gather plenty of information to characterize the atmosphere of a planet and its habitability. 
Retrievals performed considering only \uvvisnir~data would in principle allow us to determine that the planet has water vapor in the atmosphere, though the high uncertainty on the surface temperature and pressure would still make it unclear whether there could be liquid water on the surface. It would also be possible to detect oxygen, which is a key biosignature for Earth-like life, but the lack of accurate characterization of additional molecules such as \ce{CH4} and \ce{CO} to constrain the redox state of the atmosphere would not necessarily rule out the abiotic false positive case, where oxygen is generated by non-biological processes such as photolysis \citep[see, e.g.,][]{Domagal-Goldman_2014,Meadows2018}.

On the other hand, retrievals on \mir~data (\life~retrievals, in blue) could provide very strong constraints on \ce{CO2}, \ce{H2O}, and \ce{O3}, which could in principle also point to the presence of a potentially inhabited planet. However, the relationship between \ce{O2} and its photo-chemical byproduct \ce{O3} is not linear \citep[see][]{Segura2003,Kozakis2022}. Also in this case, a low-quality estimate of \ce{CH4} might hinder the vetting of \ce{O2} in a potential abiotic scenario. Nevertheless, assuming higher baseline resolution and signal-to-noise levels for \life~(see, e.g., LIFE Papers III and IX) would mitigate this problem by allowing a stronger constraint on the \ce{CH4} abundance. Earth-like abundances of \ce{N2O} would also be not possible to detect at this resolution and noise level. A more accurate and precise characterization of the thermal profile of the atmosphere, possible through \mir~retrievals, would provide more stringent results on the potential for liquid water on the surface, though no direct measurement of the surface reflectivity (and potential water and vegetation features) would be available. 

These statements only apply in the case of modern Earth analogs, which are the focus of this study. Other scenarios might behave differently, depending on the atmospheric composition. An example is the Archean Earth scenario, when \ce{CO2} and \ce{CH4} were likely much more abundant in the atmosphere. The simultaneous detection of \ce{CO2} and \ce{CH4} is believed to be a signature of life on early Earth \citep{KrissansenTotton2018}. The higher abundance of these two molecules in the early Earth would make the detection of these within reach of both \hwo~\citep[see, e.g.,][]{Young2024b,Young2024, DamianoHu2022} and~\life~\citep[see, e.g.,][i.e., \pV]{Alei2022}.
Another example would be an atmosphere with enhanced biogenic \ce{N2O}. \citet[i.e., \life~Paper XII]{Angerhausen_2024} found that this molecule could be detectable with \life-like observations if present at plausible biological production fluxes.

\subsection{Complementarity of multi-instrument observations}\label{sec:complementarity}

From what we have seen in this study, multi-instrument observations are clearly helpful for reducing biases and overall enhancing the quality of the results. For each set, the posterior distributions of the joint retrieval run are mostly at the intersection of the single-instrument results (see the corner plots in Figures \ref{appfig:validationcorner}, \ref{appfig:constanterrorbarscorner}, \ref{appfig:scaledsnrcorner}). The posteriors also show a reduced bias and a smaller variance around the true value (see the mean-squared-error bar plots in Figures \ref{fig:constanterrorbarsmse}, \ref{fig:scaledsnrmse}, and \ref{fig:validationrmse}). 

Having data spanning a larger wavelength range translates into an increase in the number of spectral features to compare the forward model against when performing a Bayesian retrieval. It further becomes easier to disentangle overlapping features, by including more bands where each specific molecule can be more unambiguously detected. This limits the range of parameter space that can accurately reproduce the data, thus shrinking uncertainties (see Figures  \ref{fig:constanterrorbarsposteriors}, \ref{fig:scaledsnrmse}, and \ref{fig:validationposteriors}). 
From the point of view of the atmospheric structure, different wavelength bandpasses sample different pressures throughout the atmosphere, allowing us to have a more precise thermal structure when including data from \hwo~and from \life, compared to the single-instrument counterparts (see Figures \ref{fig:constanterrorbarspt}, \ref{fig:scaledsnrpt}, and \ref{fig:validationpt}).

At the baseline resolution values (simplified noise and high-fidelity noise scenarios) a joint \hwolife~retrieval on a cloud-free Earth-like planet would yield very strong constraints on \ce{O2}, \ce{O3}, \ce{CO2}, and \ce{H2O},  as well as weak constraints on \ce{CH4} and \ce{CO}. In our runs, the confidence level for the detection of each of the considered species in the joint retrieval scenario is at least equal to the larger one between the two single-instrument retrievals and it increases considerably in some cases (e.g., \ce{O3}, see Figures \ref{fig:constanterrorbarsmse} and \ref{fig:scaledsnrmse}). This shows that, whenever enough information is stored in one portion of the wavelength spectrum, this is also similarly detected with a joint retrieval. If, however, both portions of the spectrum contain information on a specific parameter, then the joint retrieval performance is magnified.

The joint retrievals also slightly overestimate the ground pressure. This is probably due to the VIS+NIR spectrum which, as we have discussed in the previous subsection, is not sensitive to the atmospheric thermal profile but that has a marginally higher resolution than the \mir~thermal emission spectrum. Therefore, the forward model might favor the strong \ce{O2} lines in the reflected spectrum, which tend to skew the result towards high pressures and low abundances, as it happens in the \hwo-only retrievals. The overestimation of the ground pressure would in principle slightly increase the range of temperatures at which water can be liquid. However, the precise and accurate estimate of the surface temperature (retrieved with an uncertainty of $\approx5$ K when considering joint retrievals) would be more convincing about the habitability of a promising target, compared to the single-instrument runs.

A joint retrieval would therefore provide the highest quality of information available for a target.  While an increase in the quality of the results compared to the single-instrument retrievals may seem quite obvious since the joint retrieval can leverage all the available information throughout the wavelength range, this represents a strong validation of our base assumption that the sum of both missions would be greater than each separately. This set of runs also serves as a ``proof of concept'' for a joint characterization of a terrestrial planet since it highlights how transformational the multi-wavelength observation of a promising planet could be, especially when trying to characterize the habitability of a potential candidate host for life.

Furthermore, a detailed characterization of both the reflected and the thermal portions of the planetary spectrum would help break the radius-albedo ambiguity by enabling an energy-budget analysis of the planet. By constraining both the albedo and the abundance of various greenhouse gases, it would be possible to estimate the expected greenhouse effect given the observed surface temperature and atmospheric profile, which would be a prerequisite for an accurate assessment of the habitability of the planet. Estimating the surface temperature with high precision would also refine the search for life by determining the likelihood of the presence of surface water; this would be a key piece of information needed to identify a promising target for further in-depth study, and it would likely not be prior available information.

Finally, the population-level results will be much more robust with both reflected light and thermal emission. Observations in both wavelength ranges will help us characterize the diversity of the planetary atmospheres that both missions will observe, paving the way for an unbiased, complete picture of the demographics of the sample.

\subsection{Impact of noise}\label{sec:noise}

At this stage, accurate noise models that are specific to the two concepts are not yet fully developed. The development of these tools is strongly coupled with the architecture trades and the technological assessments that are currently in progress and will be executed over the coming years. However, it is assumed that we will be dealing with very small signals. In this ``low-quality'' data regime, different noise instances might be the reason for the detection or non-detection of specific molecules. 

In this study, we used various noise instances to explore the impact that noise has in the retrievals. As explained in Section \ref{sec:inputspectrum}, the noise level was the only difference between the ``simplified noise'' and ``PSG/LIFE\textsc{sim} noise'' retrievals. By comparing these sets of retrievals (see Section \ref{sec:results}) we find that a more complex noise modeling causes some differences in the results. The most noticeable one is the detection of \ce{O3}, which can be decisively detected thanks to the lower noise level in the UV band in the high-fidelity noise set compared to the very weak confidence in the detection in the simplified noise set. This highlights the great impact that accurate noise modeling could have in defining the requirements of future-generation missions and in prioritizing some architectures compared to others.
As we progress in the definition of the various concepts, the noise models will need to be updated and these results will likely change. It is therefore important at this stage of the concepts' maturation to benchmark results and to simulate observations through atmospheric retrievals, since these will provide information that will be fed back into the definition of the preferred architectures.

\subsection{Known limitations of this study}\label{sec:limitations}

The results discussed in this work are valid within the assumptions that were made. To simplify the problem and to provide a first-order overview, we used a simplified, cloud-free scenario that is not purely motivated by physics and is unlikely to be found in reality.  In this scenario, it is possible to disentangle the correlation between the radius and the albedo through a high-resolution characterization of the Rayleigh cross section in the \uvvisnir. However, this would most likely not be the case when clouds and hazes are present \citep[see, e.g., ][]{Feng_2018,RobinsonSalvador2023,DamianoHu2022}. A \mir~observation should be less prone to such issues (see \life~Paper IX).

Furthermore, we used a simplified retrieval framework to allow for reasonable computing time (see Section \ref{sec:updates} for details). Some further biases in the detection of the most abundant species in the atmosphere (the bulk scatterer \ce{X2} and \ce{O2}) could have been caused by \texttt{pyMultiNest} since this nested sampling algorithm has been found to undersample the edges of the prior space \citep[see Appendix D of][]{himes_2022}, which is the case for these two molecules. This bias could have especially negatively influenced the retrievals in the \uvvisnir~range, for which it was possible to constrain such molecules. Retrievals on the \mir~portion of the spectrum would not have suffered from this bias, since not enough information about these molecules is contained in this range. 

We assumed that we would perfectly know the geometry of the planet-star system, set at quadrature, and that the planet signal is fixed: rotation and atmospheric dynamics that happen on daily or seasonal timescales can be ignored. 
In reality, the phase of the system is likely to be unknown in this kind of observation and will need to be accurately modeled in retrievals. 

Finally, we do not randomize the individual spectral points according to the noise, but rather we consider the noise as an uncertainty to the theoretically simulated flux points. This has been done in previous studies (see Appendix \ref{app:comparison}), but it could lead to an overestimation of the results. In the Appendix of \pIII~we compared the results of retrievals on non-randomized and randomized flux points, showing that results on non-randomized spectra could be interpreted as average estimates of the randomized-spectra results.

These issues will be overcome and improved in future studies. Some other limitations are however much more rooted in the development of these concepts themselves and the technical challenges that stem from that. 

First of all, the noise treatment should be improved and updated as long as the iterations for various architecture trades proceed. Given the impact that noise could have on the confidence of the detection of important life-related molecules (as shown by comparing the two baseline sets of retrievals), it is mandatory to keep refining the instrumental parameters that have an impact on the noise. This will also help define the observing time required for each observation to achieve the desired \SN. Once we get a better understanding of the processes that cause the spectral shape of the noise to change with exposure time, it will be possible to simulate accurate noise instances at the required observing time, instead of scaling a template noise to the desired \SN~as it was done in this study. Retrieval benchmark studies will then need to be repeated to identify the architecture that yields the best possible results.

The Habitable Worlds Observatory will very likely include a coronagraph. This will require the presence of specific, multiple spectral bandpasses in the three wavelength ranges. We would not necessarily expect to have access to a complete spectrum of the planet in the \uvvisnir~range, like it was assumed here. Although some initial studies on the observation strategy  \citep{Young2023} and on how these bandpasses should be defined for HWO \citep{Latouf2023b,Latouf2023} were performed, these are still to be confirmed. 
Since acquisitions in the same wavelength range are limited to one bandpass at a time \citep{Young2023} and since we would deal with long exposure times (of the order of hours, or even days), it is likely that the geometry of the planet-star system would change between one acquisition and another. This would complicate the modeling and the retrieval of realistic results. 

We used a relatively large prior on the radius to allow for comparable retrieval results for both wavelength ranges of interest. However, a \life-like mission should be able to provide more stringent constraints on the radius from the search campaign (see~\pII) which could be used as prior for retrievals in both wavelength ranges and therefore potentially allow a more accurate retrieval of the planetary radius. Similarly, we could imagine that an improvement in the prior estimate of the mass (e.g., leveraging prior observations and/or the search campaign of a \mir~\life-like mission) would improve the results.

At this stage, it is entirely plausible that the two missions will not fly at the same time, but rather one before the other. The added complexity of multi-epoch observations might bring its own set of challenges, as the geometry of the observation has likely changed and phase modeling will need to be taken into account. In the context of finding more robust clues for the presence of life on an exoplanet, joint retrievals on \uvvisnir+\mir~data might not be the best strategy, as one would have to merge data at different ephemerids. It might be more efficient to use the posteriors of a retrieval performed on data from one mission as priors for retrievals performed in the other spectral range. However, as we have seen in this study, biases are present in single-instrument retrievals. Therefore, it would be more realistic to select only the results we are more confident in to feed into future retrievals. For example, one should be more confident in the radius estimate given by retrievals in the \mir, rather than the \uvvisnir. Similarly, the detection of \ce{O2} through \uvvisnir~could be fed into retrievals on \mir~data to improve the determination of the molecular weight of the atmosphere. Given the strong complementarity of these results, such strategies might play a role in maximizing the yield from these missions. Further studies are required to assess which parameters are more likely to be constrained by which concept (and wavelength range) over a variety of case scenarios that go beyond a cloud-free Earth twin.

\section{Conclusions}\label{sec:conclusions}

In this study, we performed atmospheric Bayesian retrievals on simulated data from a \hwo-like and a \life-like mission to assess the science potential of the two concepts separately and in synergy with each other for the characterization of a simplified Earth twin. 
To do this, we produced simulated observations of a cloud-free terrestrial planet both in reflected light and thermal emission. Furthermore, the retrieval routine that we developed for previous LIFE-related studies \citep[e.g.,][]{Konrad2022,Alei2022,Konrad2023,Mettler2023} has been updated to consider both \uvvisnir~and \mir~spectra.

We considered two scenarios: a simplified noise model at constant error bars considering the baseline resolutions for the various wavelength ranges of interest, and a more complicated noise model using the current noise simulators available for \life~and ``6-m LUVOIR-B'' as template for \hwo. 

We conclude that:

\begin{enumerate}
    \item Retrievals considering purely data from one of the two instruments would not provide a full characterization of the atmosphere and thermal structure of the planet. %In the idealized scenario, retrievals performed on the \hwo~case would constrain the surface temperature up to a 15 K uncertainty and would not be able to sufficiently constrain \ce{CO} and \ce{N2O}, both relevant to rule out abiotic processes (for the former) and assess metabolic activity (for the latter) \citep[see, e.g., ][]{Schwieterman2018}. \life~would be able to constrain with extreme precision the thermal profile of the exoplanet (estimating the surface pressure down to 0.5 K uncertainty), though it would not be able to constrain \ce{O2}, relying purely on its photochemical by-product \ce{O3} for assessing the habitability of the planet. 
    A retrieval on \uvvisnir~data alone strongly constrains \ce{O2} and \ce{H2O}, and retrieves an upper limit on \ce{CH4}. A retrieval on \mir~data strongly constrains \ce{CO2}, \ce{H2O}, and \ce{O3}, and also an upper limit on \ce{CH4}. Neither of the two concepts can sufficiently constrain Earth-like abundances of \ce{CO} and \ce{N2O}, both relevant to rule out abiotic processes (for the former) and assess metabolic activity (for the latter) \citep[see, e.g., ][]{Schwieterman2018}.  \life~retrievals would constrain the thermal profile of the atmosphere, while \hwo~retrievals would allow to constrain the surface reflectance (in the absence of clouds).

    \item Independent of the noise and the spectral resolution, a joint \uvvisnir+\mir~retrieval can improve the confidence level of the detection of the main potential biosignatures and bioindicators.  Retrievals on \uvvisnir+\mir~ constrain all the species that the single-instrument runs estimate, but increasing the confidence of the estimates itself: \ce{O2}, \ce{CO2}, \ce{H2O}, and \ce{O3} would be detected with decisive confidence, while \ce{CO} and \ce{CH4} could be weakly detected. Joint retrievals also improve the estimate of the surface temperature and pressure, as well as the radius. 
 \item The different noise assumptions can critically impact the results: the most realistic noise scenario (PSG/\lifesim~noise) would allow to strongly retrieve \ce{O3} on \uvvisnir~only data, while it would be only possible to get a low-significance constraint on that species in the simplified noise case. This result suggests that a higher-fidelity noise model could allow more accurate results that could drive the definition of the requirements of these missions. Further, the spectral resolution and \SN~reference values are still up for discussion and in this work we only explored a small section of the \R-\SN~parameter space.  The refinement of science requirements will be an iterative process that will continue in time as technological progress is made and as noise simulators take more and more realistic instrumental processes into account.
\end{enumerate}

This has been the first study on the possible synergies between the Habitable Worlds Observatory and the Large Interferometer for Exoplanets in characterizing habitable exoplanets, in support of the development of both by the community. Both wavelength ranges provide us with their specific set of unique information and come with specific drawbacks. Yet, the scientific yield of synergistic observations in the \uvvisnir+\mir~range has the potential of being greater than the sum of its parts. Having access to multiple spectral windows into the atmosphere of a potentially habitable planet could be transformative for the search for life in the universe.

\begin{acknowledgements} 
E. Alei’s work has been partly carried out within the framework of the NCCR PlanetS supported by the Swiss National Science Foundation under grants 51NF40\_182901 and 51NF40\_205606.  E. Alei’s research was partly supported by an appointment to the NASA Postdoctoral Program at the NASA Goddard Space Research, administered by Oak Ridge Associated Universities under contract with NASA. S. P. Quanz’s work has been carried out within the framework of the NCCR PlanetS supported by the Swiss National Science Foundation under grants 51NF40\_182901 and 51NF40\_205606. E. O. Garvin gratefully acknowledges the financial support from the Swiss National Science Foundation (SNSF) under project grant number 200020\_200399. V. Kofman is supported by the GSFC Sellers Exoplanet Environments Collaboration (SEEC) and the Exoplanets Spectroscopy Technologies (ExoSpec). P. Molli\`ere acknowledges support from the European Research Council under the European Union’s Horizon 2020 research and innovation program under grant agreement No. 832428. E. Alei acknowledges T. Birbacher, F. Dannert, S. Gaudi,  S. Mahadevan, B. Mennesson, A. Roberge, G. Villanueva, and A. Young for their expertise and useful discussions. We thank an anonymous reviewer for the helpful comments.
\\
\\
\textit{CRediT author statement}.\textit{ E. Alei}: Conceptualization, Methodology, Software, Validation, Formal Analysis, Investigation, Data Curation, Writing - Original Draft, Visualization; \textit{S. P. Quanz}: Conceptualization, Resources, Writing - Review and Editing, Supervision, Project Administration, Funding Acquisition; \textit{B. S. Konrad}: Methodology, Software, Writing -- Review and Editing; \textit{E. Garvin}: Formal Analysis, Writing -- Review and Editing; \textit{V. Kofman}: Methodology,  Writing -- Review and Editing; \textit{A. Mandell}:  Methodology, Writing -- Review and Editing; \textit{D. Angerhausen}:  Writing -- Review and Editing; \textit{P. Molli\`ere}:  Writing -- Review and Editing; \textit{M. Meyer}:  Writing -- Review and Editing; \textit{T. Robinson}:  Writing -- Review and Editing; \textit{S. Rugheimer}:  Writing -- Review and Editing.
\\
\\
\emph{Software.} This research made use of: Astropy\footnote{\url{http://www.astropy.org}}, a community-developed core Python package for Astronomy \citep{astropy:2013, astropy:2018}; Matplotlib\footnote{\url{https://matplotlib.org}} \citep{Hunter:2007}; pandas \citep{reback2020pandas}; \prt~ \citep{Molliere2019}; PSG \citep{2022fpsg.book.....V}.
\end{acknowledgements}

\bibliographystyle{aa} % style aa.bst
\bibliography{biblio} % your references Y

\appendix

\section{Scattering of terrestrial exoplanets in reflected light}\label{app:scattering}

 \prt~\citep{Molliere2019} performs radiative transfer calculations including scattering through the Feautrier method \citep{1964CR....258.3189F}. We used the Feautrier method in \pV~only considering the thermal planetary radiation and we refer to Appendix A of \pV~for details on the implementation of that process, as well as \citet{Molliere2017}. Here, we report the main equations and we discuss the additional terms that are now taken into account to calculate the scattering by direct stellar light.

The radiative transfer equation (see Eq. \ref{eq:radtrans}) depends linearly on the intensity, so planetary and stellar radiation fields can be treated in an additive way. 
 
 \begin{equation}
\mu \frac{dI}{d\tau}=-I+S.\label{eq:radtrans}
\end{equation}

In this equation, $\mu=\cos{\theta}$ where $\theta$ is the angle between a light ray and the surface normal, $\tau$ is the optical depth, $I$ is the intensity, and $S$ is the source function. 

 In the Feautrier algorithm, the intensity vectors parallel and anti-parallel to the direction of a light ray ($I_+$ and $I_-$ respectively) are defined. The radiative transfer equation is then solved for the sum and the difference of these two rays (see \pV).
Two boundary conditions need to be defined at the top of the atmosphere (TOA; corresponding to $P=0$) and at the surface ($P=P_{s}$). In the case of planetary radiation scattering only (used in \pV), the conditions are: 
 
\begin{equation}
 I_+(P=0,\mu)=0\quad \forall \mu
\end{equation}

\begin{equation}
 I_-(P=P_{s},\mu)=e_{s}\ B(T_{s})+r_{s}J^{scat}(P_{s}).\label{eq:boundary}
\end{equation}

Where $e_s$ is the surface emissivity; $r_s$ is the surface reflectance; $B(T_{s})$ is the blackbody emission of the surface; and $J^{scat}$ the average scattered intensity of the radiation that comes from the top layers. The interpretation of these conditions is that: 1) there is no planetary radiation coming from the top of the atmosphere and 2) the light going upwards from the surface is composed by the thermal emission of the surface itself (as the blackbody radiation of the surface scaled by the surface emissivity) and by a portion of the incoming planetary radiation that is reflected by the surface (as the average scattered intensity of the radiation that comes from the top layers scaled by the surface reflectance). $J^{scat}$ is calculated as follows:

\begin{equation}
 J^{scat} (P_{s})= \int_0^1 I_+(P_{s})d\mu \label{jscat}
\end{equation}

In the case of scattering of stellar radiation, it is possible to split the incoming stellar intensity into two components: the incoming stellar radiation $I_*^{dir}$  along the TOA incidence angle that survives down to a given layer of interest and which gets attenuated through absorption and scattering, and the stellar light that is scattered (potentially multiple times) in the atmosphere $I_*^{scat}$.

\begin{equation}
    I_*=I_*^{dir}+I_*^{scat}
\end{equation}

The radiative transfer equation for these components becomes ($\epsilon$ being the photon destruction probability):

\begin{equation}
    \frac{dI_*^{dir}}{d\tau}+\frac{dI_*^{scat}}{d\tau} = -I_*^{dir}-I_*^{scat}+(1-\epsilon)(J_*^{dir}+J_*^{scat})
\end{equation}

Which can be split into two equations:

\begin{equation}
    \frac{dI_*^{dir}}{d\tau}= -I_*^{dir} \label{eq:direct}
\end{equation}

\begin{equation}
   \frac{dI_*^{scat}}{d\tau} = -I_*^{scat}+(1-\epsilon)(J_*^{dir}+J_*^{scat})\label{eq:scat}
\end{equation}

Equation \ref{eq:direct} is the regular attenuation of the direct light while crossing the atmosphere. Its solution is:

\begin{equation}
    I_*^{dir} =I_*^{dir}(0,\mu)\ e^{-\tau/\mu}
\end{equation}

Where $I_*^{dir}(0,\mu)$ is the irradiation at the TOA. 

For Eq. \ref{eq:scat}, the calculation of $J_*^{dir}$ is dependent on the geometry of the problem.

In the case of isotropic stellar irradiation,  $I_*^{dir}(0,\mu)$ is integrated over all possible angles  ($-90^\circ$ to $+90^\circ$ angle between the ray and the direction normal to the surface, which corresponds to the light that comes from the top). This translates as:

\begin{equation}
    I_*^{dir} (0) =\begin{cases} F_*^{dir}(0)/\pi, & \mbox{if } \mu < 0\\ 0, & \mbox{otherwise} \end{cases}
\end{equation}

Where $F_*^{dir}(0)$ is the stellar flux scaled at the distance of the planet. The average direct intensity over all angles is then:
\begin{equation}
    J_*^{dir}= \frac{1}{2}\int_{-1}^1 I_*^{dir}d\mu= \int_{0}^1 I_*^{dir}d\mu\label{jdir}
\end{equation}

For non-isotropic irradiation (i.e., the stellar irradiation is incident under an angle $\theta_*$ of cosine $\mu_*$), the value of $F_*^{dir}(0)$ is:

\begin{equation}
    F_*^{dir}(0) = \int_{\Delta\Omega_*} I_*^{dir}\mu d\Omega \approx I_*^{dir}(0)\mu_* \Delta\Omega_*
\end{equation}

Where $\Delta\Omega_*$ is the solid angle subtended by the stellar disk from the atmospheric location. In this case, $J_*^{dir}$ is:

\begin{equation}
    J_*^{dir}=\frac{I_*^{dir}\Delta\Omega_*}{4\pi} = \frac{F_*^{dir}(0)\ e^{-\tau/\mu_*}\Delta\Omega_*}{4\pi\mu_*}
\end{equation}

The stellar light that can be scattered at the surface is then calculated analogously to Eq. \ref{jscat} for the stellar radiation field and added to the planetary radiation field.  $J^{scat}$ will now automatically include both contributions from the atmospheric thermal radiation, and the stellar one, scattered at least once inside the atmosphere.

To meet the boundary condition, the mean direct radiation must be included in the term that takes into account the portion of radiation reflected by the surface. Therefore, Equation \ref{eq:boundary} becomes

\begin{equation}
    I_-(P_{surf}) = e_{s}\ B(T_{surf})+r_{s}\left[J^{scat}(P_{surf})+ J_*^{dir}(P_{surf})\right]
\end{equation}

In our study, we simulate the quadrature phase of the planet by assuming a non-isotropic scenario with a specific angle $\theta_*=77.756$ degrees. In principle, every region of the planet would have a specific angle of incidence of the stellar light, so we should take into account the reflected spectrum of each pixel encompassed by the field of view. In the case of a Lambertian surface, which is what we assume in this study, this calculation can be simplified by averaging the angles across the field of view. This produces an ``average'' incident direct light angle of 77.756 degrees, which allows us to run a single radiative transfer simulation with reasonable results. For more information on this treatment, used widely in the Planet Spectrum Generator suite, we refer to the PSG Handbook \citep{2022fpsg.book.....V}. 

The surface reflectance $r_s$ assumed in the scattering calculation is a free parameter in our runs. It represents the specific reflectance of the surface material. In principle, \prt~can assume a wavelength-dependent reflectance of a custom surface. The user can take advantage of known reflectivity databases \citep[e.g., the United States Geological Survey Spectral Library, ][]{Kokaly2017} to fill this variable. In this study, we assume a wavelength-independent value of $r_s$ of 0.1 (an average value for habitable water-rich terrestrial planets) both when producing the input spectrum and when retrieving the $r_s$ estimate. 

The value of $r_s$ is therefore different from what is commonly assumed to be the albedo of the planet as a whole, as no contribution from the atmosphere is considered aside from the extinction of the incoming ray that is reflected by the surface. However, in the case of thin enough atmospheres with no clouds such as the one considered in this study, the majority of the scattered light that would be detected originates from the surface. For this reason, in this specific case, $r_s$ can be considered a proxy for planetary albedo.
The Bond albedo of the planet \textit{a posteriori} can be retrieved from the planetary radius estimate and the stellar luminosity, as it was done in \citet{Konrad2023,Mettler2023}.

\section{Idealized high-resolution low-noise scenario}\label{sec:photonnoise}

\begin{figure*}
     \centering 
     \textbf{Idealized high-resolution low-noise scenario}\par\medskip

\begin{subfigure}[b]{0.45\textwidth}
         \centering
         \includegraphics[width=\textwidth]{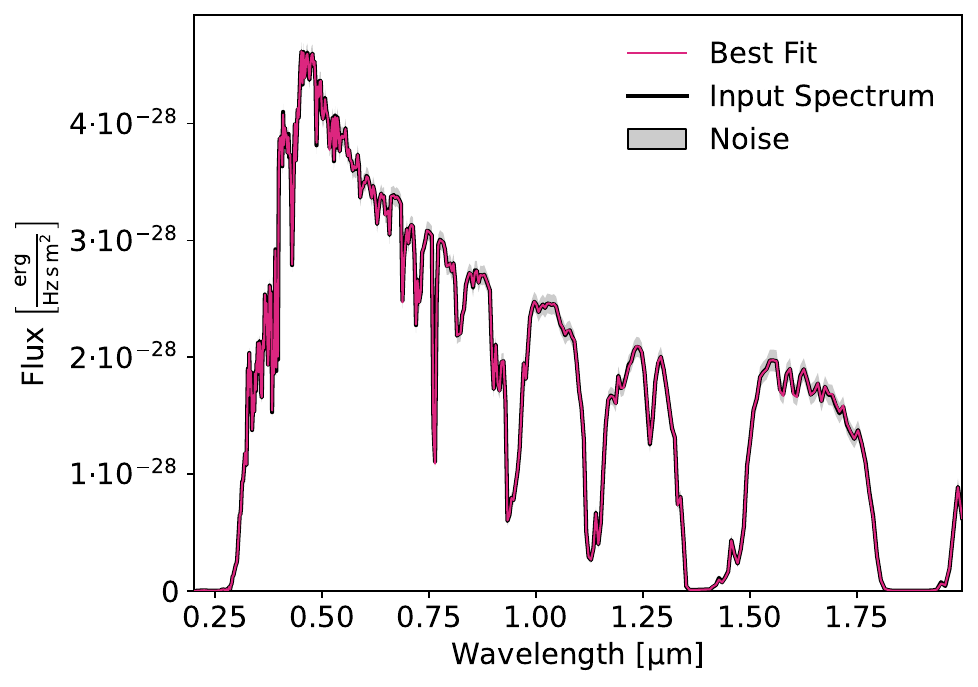}
     \end{subfigure}
          \begin{subfigure}[b]{0.45\textwidth}
         \centering
         \includegraphics[width=\textwidth]{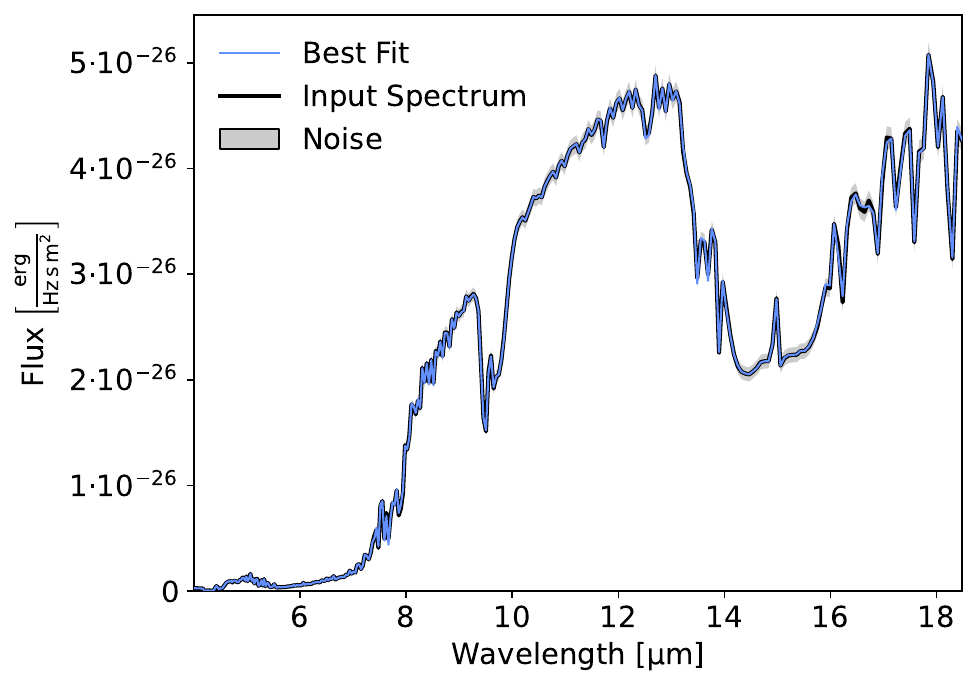}
     \end{subfigure}

     \begin{subfigure}[b]{0.92\textwidth}
         \centering
             \hspace{-0.4cm} \includegraphics[width=\textwidth]{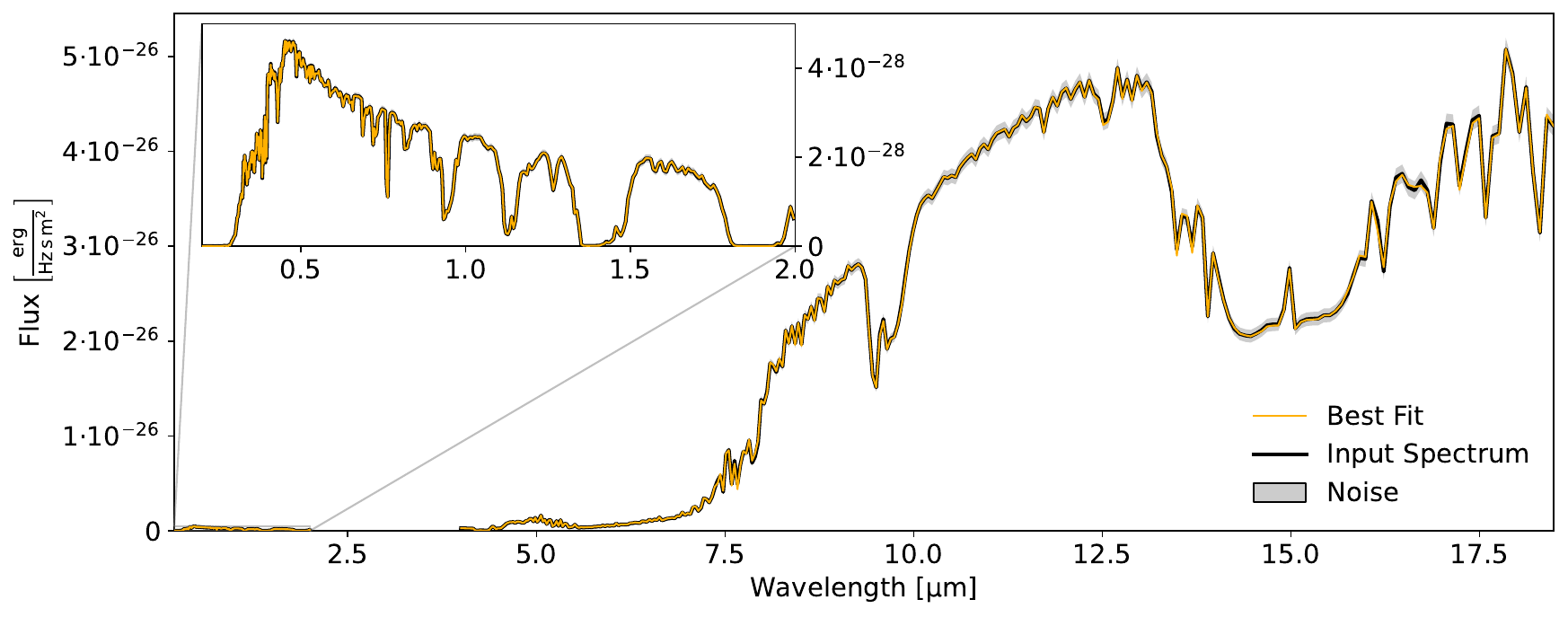}

     \end{subfigure}
          \begin{subfigure}[b]{0.5\textwidth}
         \centering
         \includegraphics[width=\textwidth]{Figures/legend.pdf}

     \end{subfigure}
        \caption{Best-fit spectra retrieved in the first retrieval set (photon noise only at \SNv{50}, \Rv{1000}) for the three scenarios: \emph{Left panel}: pure \hwo~retrieval (magenta);   \emph{Central panel}: pure \life~retrieval (blue); \emph{Right panel}: \hwolife~retrieval (yellow). In all panels, the best-fit spectra are compared with the input spectra (black lines) with error bars (gray-shaded areas).}
        \label{fig:validationspectrum}
\end{figure*}

\begin{figure*}
     \centering
     \textbf{Idealized high-resolution low-noise scenario}\par\medskip
          \begin{subfigure}[b]{0.33\textwidth}
         \centering        \includegraphics[width=\textwidth]{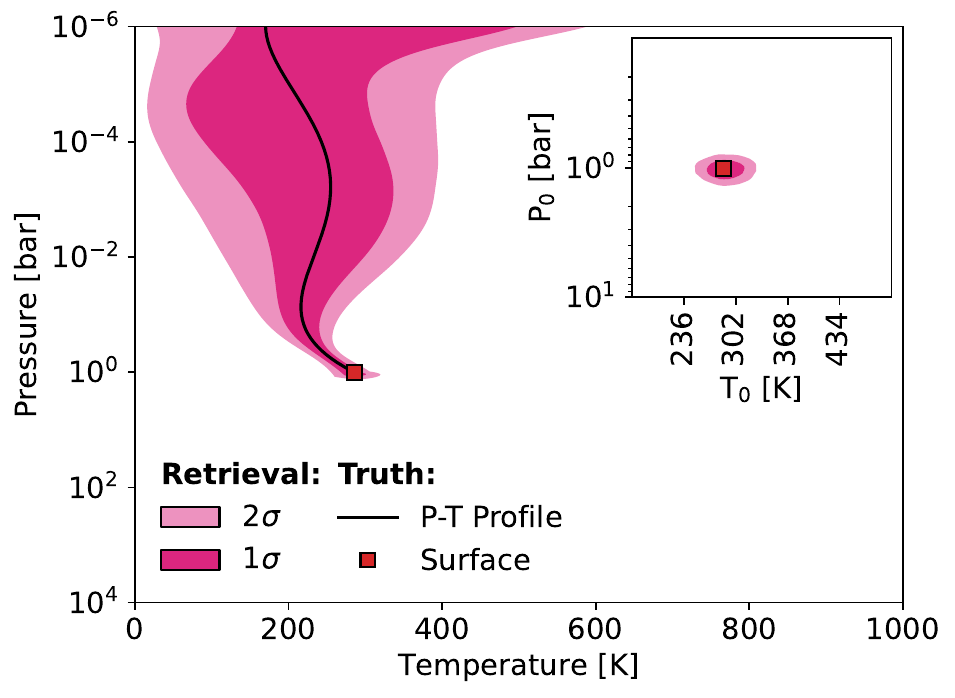}
     \end{subfigure}
     \hfill
     \begin{subfigure}[b]{0.33\textwidth}
         \centering
         \includegraphics[width=\textwidth]{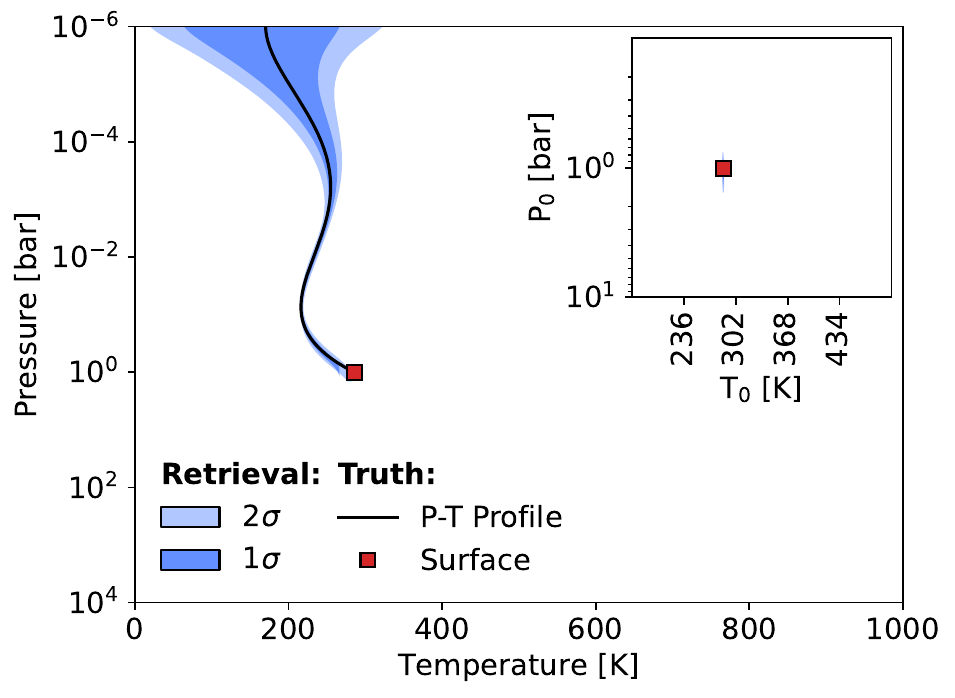}
     \end{subfigure}
\hfill
     \begin{subfigure}[b]{0.33\textwidth}
         \centering
         \includegraphics[width=\textwidth]{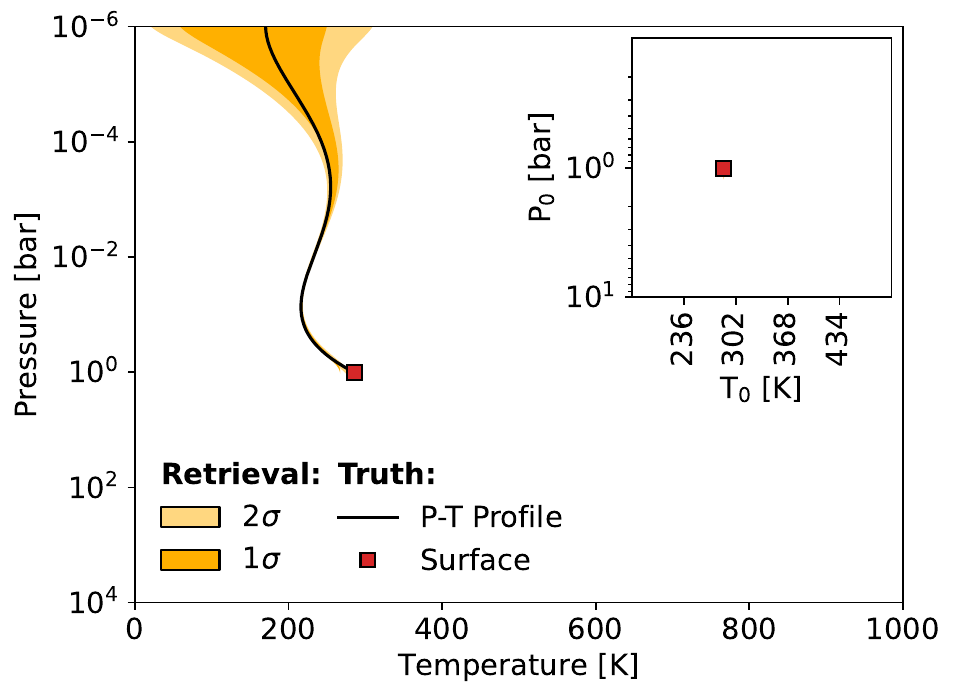}
     \end{subfigure}
          \begin{subfigure}[b]{0.5\textwidth}
         \centering
         \includegraphics[width=\textwidth]{Figures/legend.pdf}

     \end{subfigure}
        \caption{Retrieved pressure-temperature profiles for the first retrieval set (photon noise only at \SNv{50}, \Rv{1000}): \emph{Left panel}: pure \hwo~retrieval (magenta);   \emph{Central panel}: pure \life~retrieval (blue); \emph{Right panel}: \hwolife~retrieval (yellow). In all panels, the 2-$\sigma$ and the 1-$\sigma$ intervals are shown in increasingly darker hues, as well as the input profiles (black lines) for comparison. Inside each panel, the inset plot shows the 2D posterior space of the ground pressure and temperature. In all panels and the inset plots, the surface pressure and temperature point in the \pt~space is shown as a red square marker.}
        \label{fig:validationpt}
\end{figure*}

\begin{figure}
\centering
 \textbf{Idealized high-resolution low-noise scenario}\par\medskip
 \includegraphics[width=\linewidth]{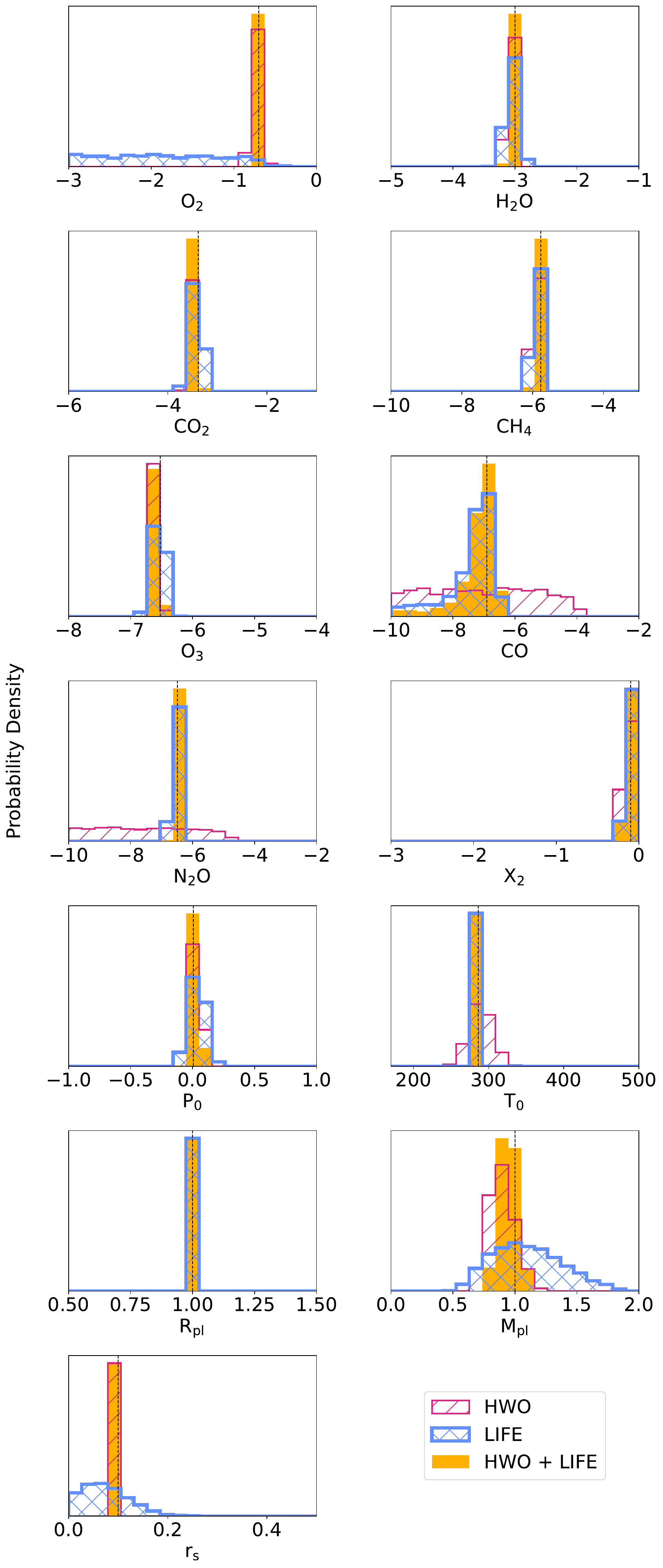}

\caption{Posterior density distributions from the first set of retrievals (photon noise only at \SNv{50}, \Rv{1000}). The black lines indicate the expected values for every parameter. \hwo~posteriors are shown in magenta with diagonal hatching; \life~posteriors are shown in cyan with crossed hatching; \hwolife~posteriors are shown as fully colored gold histograms. }\label{fig:validationposteriors}
\end{figure}

\begin{figure}
     \centering
               \textbf{Idealized high-resolution low-noise scenario}\par\medskip
     \begin{subfigure}[b]{\linewidth}
         \centering
         \includegraphics[width=0.9\textwidth]{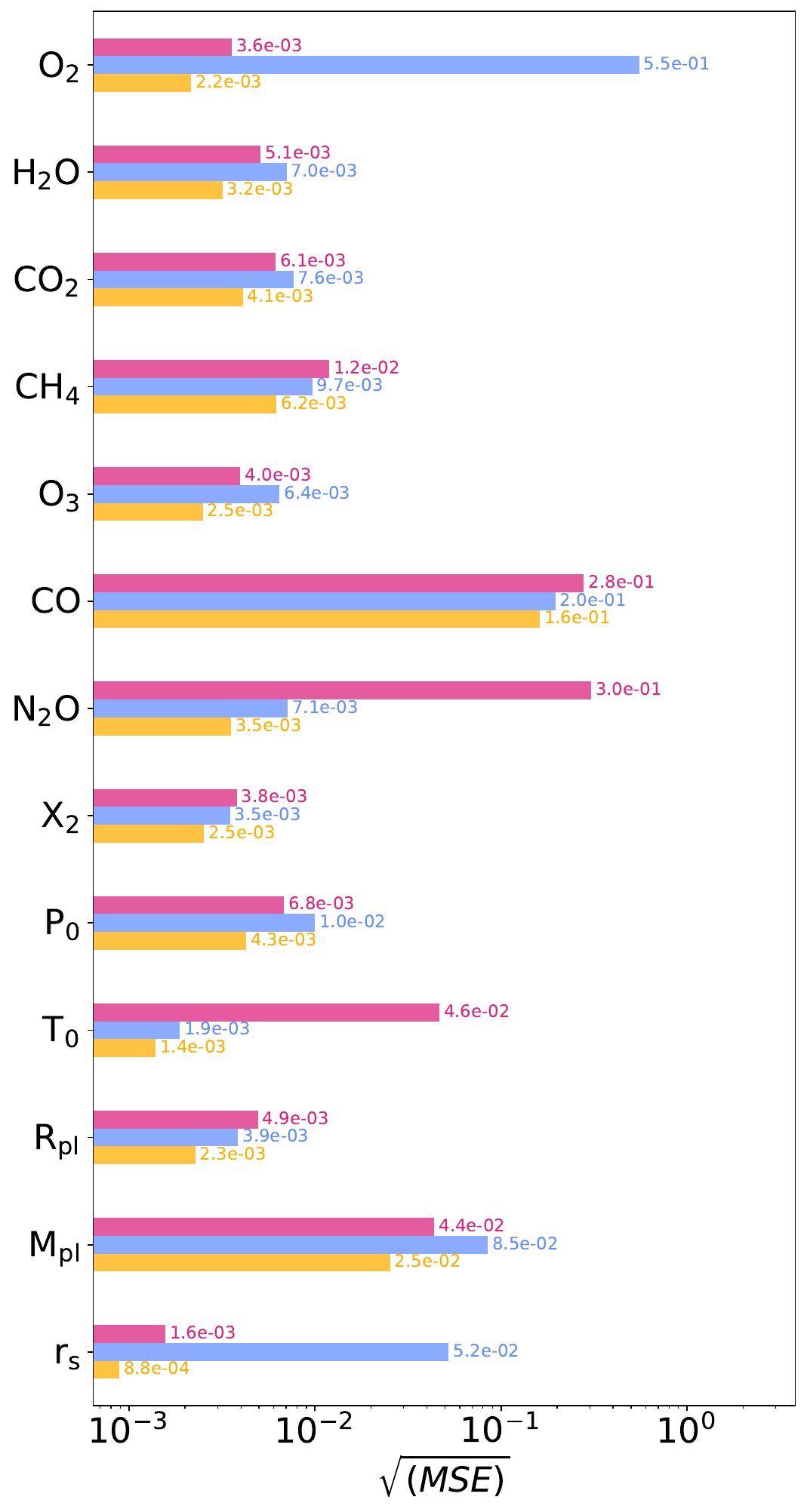}
     \end{subfigure}

          \begin{subfigure}[b]{0.5\textwidth}
         \centering
         \includegraphics[width=\textwidth]{Figures/legend.pdf}

     \end{subfigure}
        \caption{Square root of the mean squared error (see Equation \ref{eq:mse}) for relevant parameters in the first set of retrievals (photon noise only at \SNv{50}, \Rv{1000}). }
        \label{fig:validationrmse}
\end{figure}

\begin{figure*}[h]
     \centering
     \textbf{Idealized high-resolution low-noise scenario}\par\medskip

     \begin{subfigure}[b]{0.9\textwidth}
         \centering
         \includegraphics[width=\textwidth]{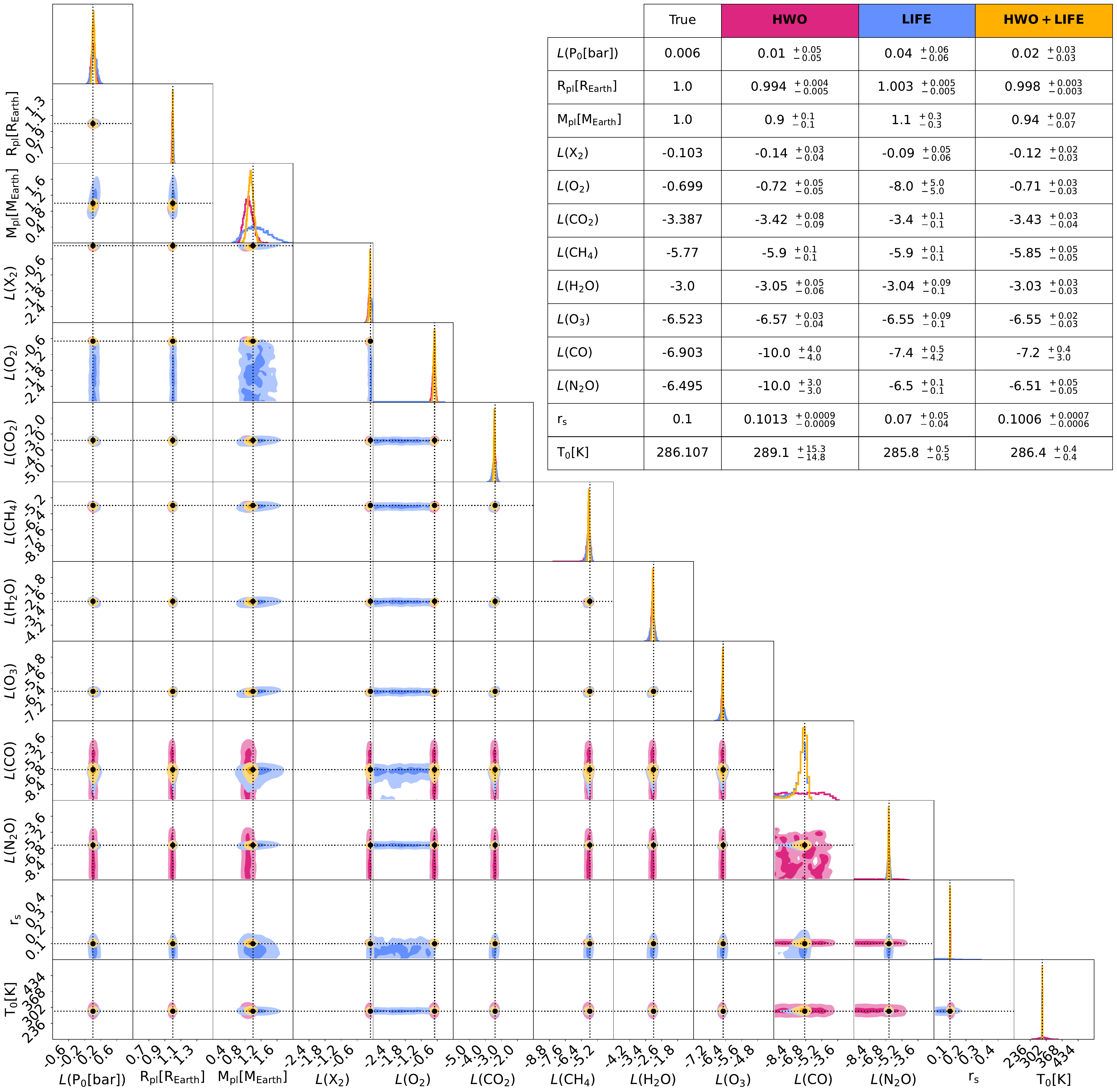}
     \end{subfigure}
        \caption{Corner plot for the posterior distributions from the first set of retrievals (photon noise only at \SNv{50}, \Rv{1000}). The black
lines indicate the expected values for every parameter. The median and 1-$\sigma$ uncertainties for relevant retrieved and derived parameters are shown in the table in the top right corner. The scenarios are color-coded according to Table \ref{tab:runcategories}.}
        \label{appfig:validationcorner}
\end{figure*}

In this set of runs, we simulated the spectrum of a cloud-free Earth at \Rv{1000} with uncertainties based on a \SNv{50} at \mic{0.55} and \mic{11.2}, only taking photon noise into account. This serves as a validation of the updated retrieval routine and represents an idealistic best-case scenario. 
In Figures \ref{fig:validationspectrum}, \ref{fig:validationpt}, and \ref{fig:validationposteriors}, we show the best fit of the spectrum compared to the input data,  the retrieved \pt~profiles, and the posteriors of the main physical and chemical parameters for the first retrieval set. The corresponding corner plot is shown as Figure \ref{appfig:validationcorner}. The square root of the mean squared error for various parameters is shown in Figure \ref{fig:validationrmse}.

As can be noticed in Figure \ref{fig:validationspectrum}, we are able to accurately and precisely reproduce the input spectra in all the scenarios. The 1-, 2-, and 3-$\sigma$ envelopes of the retrieved spectrum lie well within the photon noise uncertainty. For better clarity of this set of plots, we do not show the $\sigma$ envelopes for the best-retrieved spectra in this case, but only the best fit. %To show it more clearly, we calculated the residual as the difference (in percent) between the true and the retrieved spectrum (see Figure \ref{fig:constanterrorbarsresspectrum}). For all cases in this scenario, the uncertainty of the best fit is well below the assumed photon noise.

Retrieving an accurate \pt~profile is satisfactory with \life: the profile is correctly determined down to pressures of approximately $10^{-3}$ bar. This is expected, since for a cloud-free thin-atmosphere scenario the majority of the contribution to the thermal emission spectrum is coming from the lower, denser layers of the atmosphere.  In the \life~retrieval, the surface pressure and temperature are also retrieved with high precision and accuracy (with a 1-$\sigma$ uncertainty of 0.06 dex for the pressure, and only 0.5~K for the temperature). The \hwo~retrieval is less accurate, as can be seen from the larger envelopes in the \pt~profile. The retrieval of surface temperature and pressure is relatively accurate (the 1-$\sigma$ interval uncertainty of \Ts~is around 15~K in this case).  
The joint retrieval can estimate the \pt~profile with better accuracy and precision than the single-instrument retrievals. Figure \ref{fig:validationposteriors} shows that, in the case of the idealized \hwolife~retrieval, the uncertainties on the surface pressure and temperature are reduced (up to a 1-$\sigma$ uncertainty of 0.03 dex for \Ps~and 0.3 K for \Ts). This result stems from the high quality of the spectrum, which allows the retrieval framework to correctly model pressure and thermal broadening of each line -- and especially the deeper layers of the atmosphere that contribute the most to the spectrum.

The retrieval of the radius is very accurate and precise in all three retrievals: the estimate of the radius and corresponding 1-$\sigma$ uncertainty is $1\pm0.01\ R_\oplus$ or lower for all scenarios. Retrievals that include \mir~data (\life~and \hwolife~retrievals) can leverage the thermal emission of the planet, which provides a strong constraint on the radius of the planet (see, e.g., \pIII). On the other hand, the result of the \hwo~retrieval is dependent on the assumptions made in this study. Since we neglect clouds and assume a perfect knowledge of the geometry of the system, the radius/albedo degeneracy in the \uvvisnir~can be easily disentangled. This will likely not be the case in reality, as discussed in Section~\ref{sec:limitations}.

Concerning the planetary mass, even the most idealized scenario of a \mir~\life-like retrieval only slightly further constrains the mass of the planet compared to the prior (retrieving a mass of $M_{pl}=1.08\pm0.30\ M_\oplus$). This happens because of the well-known degeneracy between gravity and chemical abundances, which contribute to the scale height of the atmosphere (see \pIIIaV~for details). On the other hand, the \uvvisnir~retrieval allows for a more precise estimate of the mass with an uncertainty of about 10\%, but a less accurate result (estimating $M_{pl}=0.88\pm0.10\ M_\oplus$). Considering both wavelength ranges together would instead improve the mass estimate to $M_{pl}=0.94\pm0.07\ M_\oplus$, with the true value within the 1-$\sigma$ confidence interval. 

The surface reflectance is accurately and precisely determined by the \hwo~retrieval. On the other hand, the \life~retrieval can only rule out surface reflectances higher than 0.3-0.4. Further, there are slight correlations between  $r_s$ and the major atmospheric absorbers in the \mir. The combined \hwolife~retrieval retrieved this parameter correctly. The surface reflectance is also impacted, like the radius, by the assumptions we make in these studies regarding geometry and clouds (see Section \ref{sec:limitations}).

At such high spectral resolution and signal-to-noise ratio, the retrievals are generally precise and accurate at retrieving the abundances of the main atmospheric species. There are, however, a few exceptions. Even in the most idealized \hwo~retrieval, it is not possible to have a detection of \ce{CO} and \ce{N2O} (only upper limits are retrieved). These two molecules do not have significant lines in this wavelength range (the opacity being 3-4 orders of magnitude fainter than the main absorbers in this range, such as \ce{O2} and \ce{H2O}). 
In the idealized \life~retrieval, \ce{O2} cannot be sufficiently constrained, because of the lack of significant lines in the \mir~wavelength range.  \ce{CO} has a strong detection peak around the true value, but lower abundances cannot be entirely ruled out. This is due to its main spectral features being at short \mir~wavelengths ($\leq$\mic{4.5}), where the noise is large due to the lower thermal emission of the planet. These spectral lines also overlap with features of more abundant species (mainly \ce{H2O}, \ce{CO2}, and \ce{O3}) which makes the less abundant species harder to detect. 

When retrieving over the entire wavelength range, not only are all the species correctly retrieved, but there is a considerable increase in the precision of the retrieved estimate (a factor 2-10 on the 1-$\sigma$ uncertainty between a retrieval performed on only one portion of the spectrum, compared to the joint retrieval, see Figure \ref{fig:validationrmse}).

\section{Comparison with previous works}\label{app:comparison}

While the current study is, to our knowledge, the first retrieval study that considers the joint performance of a direct imaging \uvvisnir~mission (HWO-like) together with a \mir~mission (\life-like), many retrieval studies in the last few years have tackled the potential of future-generation instruments in characterizing Earth-like planets. We deem it useful to compare the single-instrument retrievals we performed against previous studies in the field.

\subsection{\hwo-like simulations}
From the \uvvisnir~point of view, none of the previous studies can be directly compared with ours because of strong differences in the setup. However, we will discuss similarities between our results and the ones in \citet{Feng_2018}, \citet{DamianoHu2022}, \citet{RobinsonSalvador2023}, and \citet{Young2023}. 

The work performed in \citet{Feng_2018} concerned a study on the preferred resolution and \SN\ to allow the characterization of an Earth twin. The authors considered an albedo model coupled with a direct-imaging noise simulator to provide the observed planet-to-star flux ratio. Their retrieval framework included cloud parameters, at the expense of a simplified atmosphere parameterization (isothermal \pt~profile with only \ce{H2O}, \ce{O2}, \ce{O3} considered in the retrieval). The parameter estimation module used Markov Chain Monte Carlo instead of Nested Sampling.
Furthermore, in \citet{Feng_2018} only a wavelength range between \mic{0.4-1.0} was considered, at resolutions 70 and 140.
Like in our third set of retrievals, the \SN\ was wavelength-dependent and set to a reference value of 5, 10, 15, 20 at \mic{0.55}. To compare with our results, we will focus on the \SNv{10} and \Rv{140} results, which match the quality of the VIS band in this work. 

In \citet{Feng_2018} the correlation between gravity, pressure, and atmospheric abundances is also present. However, the authors notice a strong radius-albedo degeneracy, which is not noticeable in our case given the absence of clouds. Gravity and mass remained unconstrained compared to the priors in \citet{Feng_2018}, which is consistent with our results. In both studies \ce{H2O}, \ce{O3}, and \ce{O2} are detected successfully at this resolution and \SN, though wider uncertainties are present in \citet{Feng_2018} compared to our study. This can be explained by the larger information content that we exploit in our retrieval, since we consider both UV and NIR wavelengths, while \citet{Feng_2018} is focused on the VIS range. 

\citet{DamianoHu2022} performed retrievals using a reflected light forward model and considering \ce{H2O}, \ce{CO2}, \ce{O2}, \ce{O3}, and \ce{CH4} as free parameters, assuming \ce{N2} as a filler gas. In their work, the atmosphere is assumed to be isothermal and clouds were also taken into account. The authors considered a variety of scenarios: an Earth-like atmosphere, an early Earth, a \ce{CO2}-dominated atmosphere with \ce{O2} and clouds, and a dry \ce{CO2}-dominated atmosphere. For every scenario, they focused their attention on the information content of the NIR band to correctly characterize the atmosphere. They explore various resolutions and \SN\ values. The authors retrieve their error bars by fixing the \SN\ at the reference wavelength bin corresponding to the maximum value of the spectrum, which may vary depending on the scenarios. 
To compare our results with this work, we focus on the modern Earth scenario with \Rv{140/40} for VIS and NIR and \SNv{10}, which is the closest setup to the one we simulate. These retrievals correctly identify the main components of the atmosphere (\ce{N2}, \ce{O2}, \ce{H2O}, \ce{O3}) and do not constrain \ce{CO2} and \ce{CH4}, which is consistent with our results. According to the authors, to have an upper limit on the methane abundance, it would be necessary to increase the \SN\ to 20. Despite the differences in the parameterization of the \pt~profile, also in \citet{DamianoHu2022} the ground pressure is found to be overestimated.

A more recent study by \citet{RobinsonSalvador2023} explored the detectability of an Earth twin through HabEx/LUVOIR varying the wavelength range and \SN\ in each band (considering three cases: optical only at \SNv{35}, \uvvisnir~at \SNv{20}, and VIS/NIR at \SNv{10/45}). The authors used a single-scene albedo model for the reflected light spectrum and a constant error bar \SN\ whose reference points were at \mic{0.4} for the UV, \mic{0.55} for the VIS, and \mic{1} in the NIR. 
Their results on the VIS retrieval at \SNv{35} were consistent with higher \SN cases in \citet{Feng_2018}. 
As for our comparison, we discuss our results (specifically, the \hwo~retrieval in the simplified noise scenario) with the \citet{RobinsonSalvador2023} retrieval on the whole \uvvisnir~range, which is the most similar setup despite a factor 2 difference in the \SN. The authors successfully retrieved \ce{O2}, \ce{H2O}, \ce{O3}, and only retrieve upper limits for \ce{CO2} and \ce{CH4}, in line with our results. By comparing our results with Figure 10 in \citet{RobinsonSalvador2023} we can observe similar variances in the posteriors of the constrained species, except for the \ce{O3} posterior. In our model, we cannot rule out abundances lower than $10^{-7}$, while the species is constrained in the earlier study. This is probably a result to be expected when increasing the \SN\ to 20, though such retrieval was not performed in this work. 

Finally, we compare our results with a recent publication by \citet{Young2023}. Here, the authors used atmospheric retrievals to identify the best observing strategy for the Habitable Worlds Observatory coronagraph. They performed retrievals on various combinations of wavelength bandpass regions, defining a decision tree that would allow us to differentiate between a modern Earth-like planet and an Archean Earth-like planet. They used the same retrieval framework as \citet{RobinsonSalvador2023} with a nominal \SNv{10} and the current baseline resolutions for the \uvvisnir~range (\Rv{7/140/70}).  When performing retrievals for all the considered bandpasses (three in total, centered at \mic{0.75}, \mic{0.85}, and \mic{1.65} with a 20\% bandpass width each), \citet{Young2023} retrieve \ce{H2O}, \ce{O2}, and get upper limits on \ce{CO2} and \ce{CH4}. This is consistent with what we find in the \hwo~retrievals. On the other hand, we find better constraints on the albedo and radius of the planet compared to \citet{Young2023}. This is to be expected considering that our retrievals can leverage the full \uvvisnir~spectrum and neglect clouds, which have been known to impact these parameters. \citet{Young2023} also considered different priors on these parameters compared to our studies, which can explain the differences in the results. The mass is unconstrained (compared to the chosen prior) in both studies, and the ground temperature and pressure show similar uncertainties.

\subsection{\life-like simulations}

Regarding the \mir~portion of the spectrum, we compare our results with \citet{Konrad2022} i.e., \pIII. The forward model and the retrieval routine used in \pIII~is overall consistent with the one used in this study, apart from the updates mentioned in \citet{Alei2022} (\pV) and Section \ref{sec:updates} of this manuscript.  
In \citet{Konrad2022}, a grid of spectral resolutions, \SN\ values, and wavelength ranges were explored. The \SN\ was calculated through \lifesim, in a very similar way as described in Section \ref{sec:inputspectrum}. To compare with the present work, we focus on the retrieval performed in \pIII\ considering \Rv{50}, \SNv{10}, and 
\mic{4-18.5}. We find very similar results between this model and the \life~retrieval in the third set of models (PSG/\lifesim~noise). \ce{N2}, \ce{O2}, \ce{CO}, \ce{N2O} are unconstrained in both models. On the other hand, \ce{CO2}, \ce{O3} and \ce{H2O} are constrained though the mean of the posterior in our set of models is slightly shifted towards higher pressures by about 0.5 dex, a result of the slightly underestimated ground pressure that we observe in our run. In both cases, an upper limit on methane can be retrieved, though it is impossible to rule out abundances below $10^{-6}$. This happens because the methane abundance is right at the sensitivity limit of the instrument at this resolution and \SN\ \citep[see][for more details]{Konrad2022}.

The setup of the framework validation is also consistent with the first grid of results shown in this manuscript (i.e., \Rv{1000}, \SNv{50} only considering photon noise). The precise and accurate \pt~profile retrieval in our model is consistent with the previous work, as are the estimates of the major components of the atmosphere. As expected in the \mir, the validation run in \citet{Konrad2022} also does not detect \ce{O2} nor significantly reduces the estimate of the mass of the planet.

%Overall, we conclude that the results are compatible with what was found in previous studies. 

\section{Ancillary Plots and Tables}

\begin{figure*}[b]
     \centering
               \textbf{Simplified Noise}\par\medskip

     \begin{subfigure}[b]{0.9\textwidth}
         \centering
         \includegraphics[width=\textwidth]{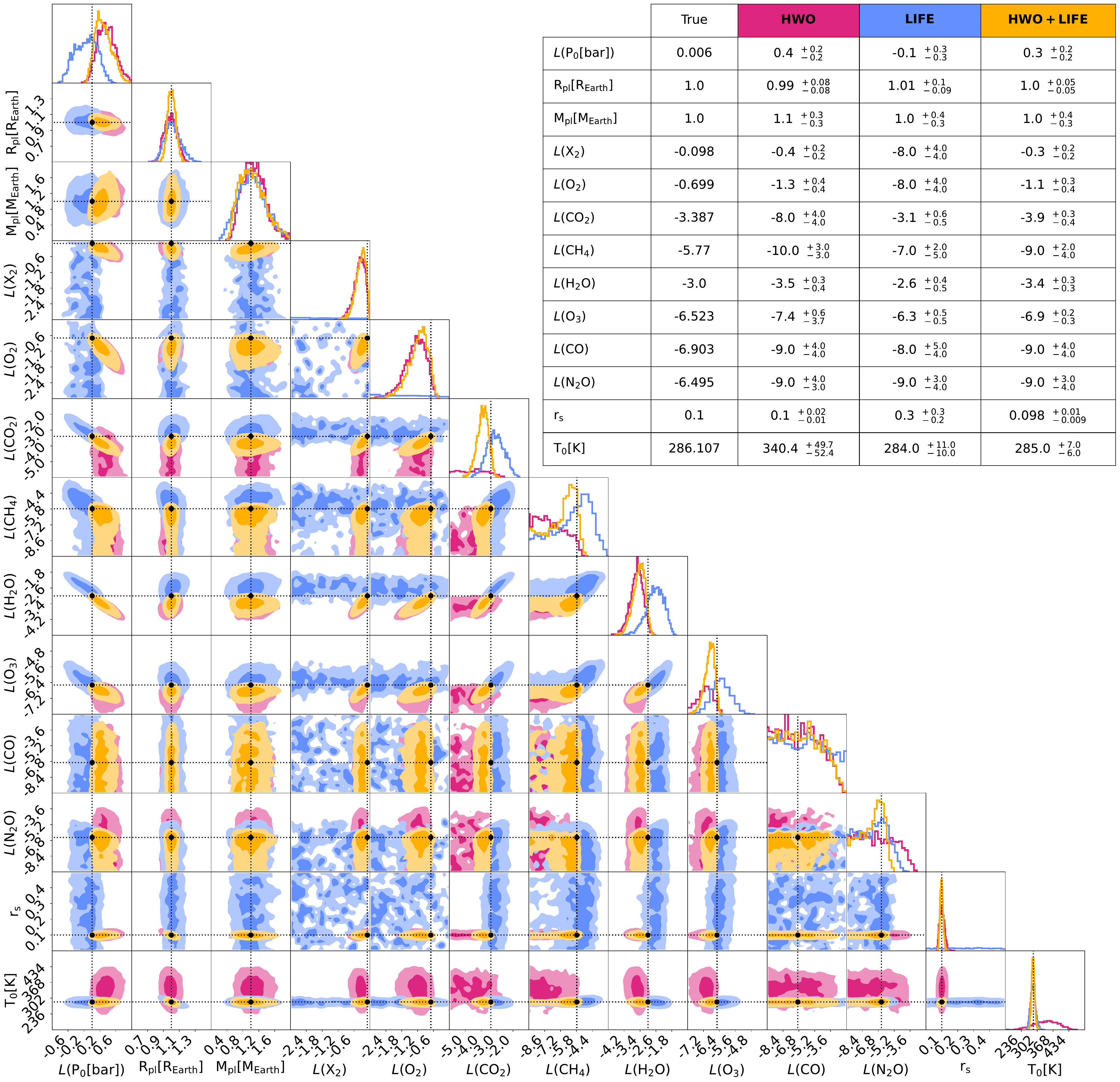}
     \end{subfigure}
        \caption{Corner plot for the posterior distributions from the second set of retrievals (simplified noise). The black
lines indicate the expected values for every parameter. The median and 1-$\sigma$ uncertainties for relevant retrieved and derived parameters are shown in the table in the top right corner. The scenarios are color-coded according to Table \ref{tab:runcategories}.}
        \label{appfig:constanterrorbarscorner}
\end{figure*}

\begin{figure*}[b]
     \centering
               \textbf{PSG/\lifesim~Noise}\par\medskip

     \begin{subfigure}[b]{0.9\textwidth}
         \centering
         \includegraphics[width=\textwidth]{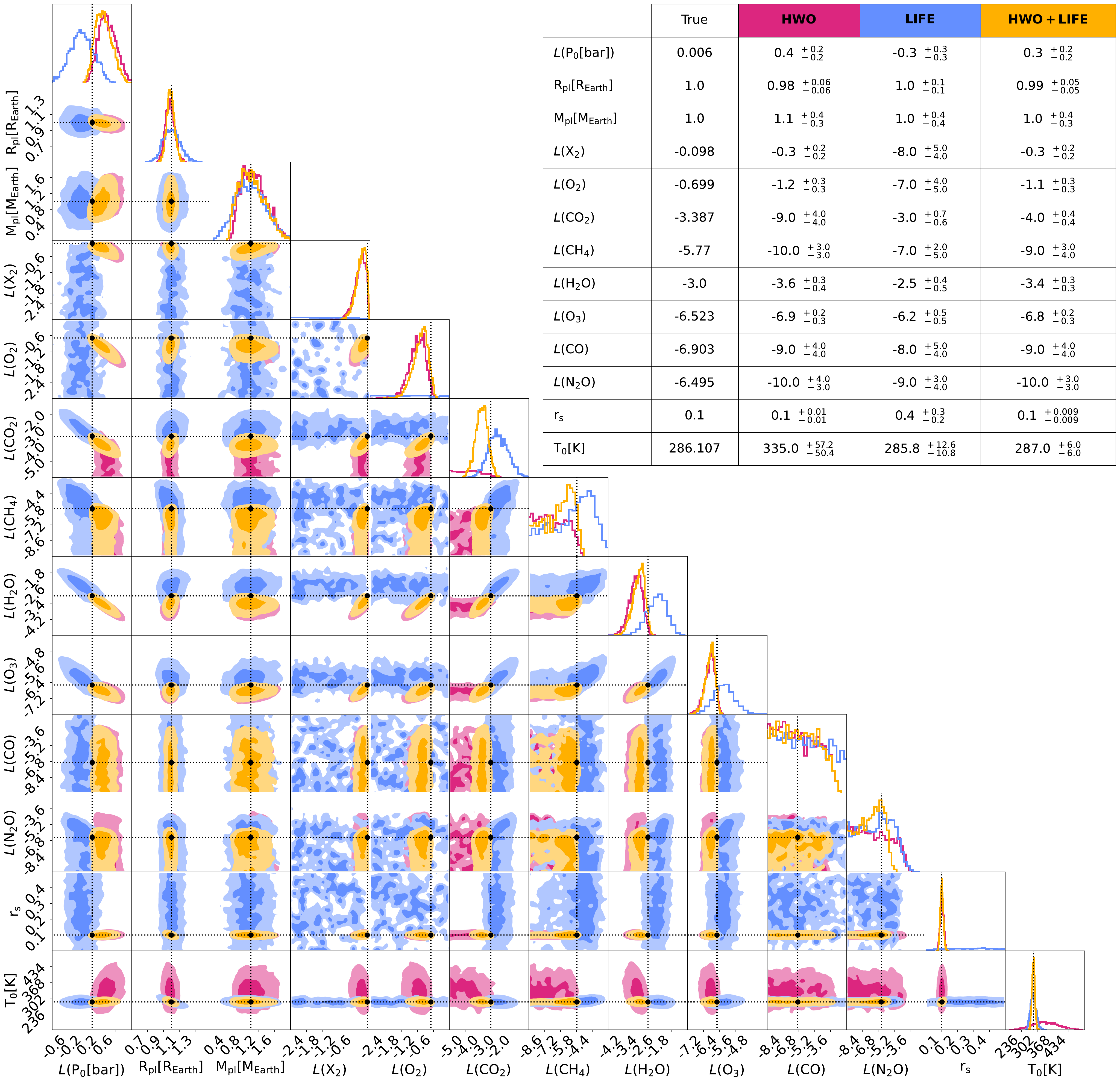}
     \end{subfigure}

        \caption{Corner plot for the posterior distributions from the third set of retrievals (PSG/\lifesim~noise). The black
lines indicate the expected values for every parameter. The median and 1-$\sigma$ uncertainties for relevant retrieved and derived parameters are shown in the table in the top right corner. The scenarios are color-coded according to Table \ref{tab:runcategories}.}
        \label{appfig:scaledsnrcorner}
\end{figure*}

\begin{table*}
\caption{Summary of the parameters used in the retrievals, their expected values, and their prior distributions. } 
\label{table:parameters} 
\centering 

        \setlength\extrarowheight{1pt} 
 \begin{tabular}{lllc}
    \hline\hline
    Parameter & Description & Priors & Input values \\ % table heading

    \hline
    $a_4$ & \pt~Parameter (Degree 4) & $\mathcal{U}(0,10)$ & 1.674 \\
    $a_3$ & \pt~Parameter (Degree 3) & $\mathcal{U}(0,100)$ & 23.120 \\
    $a_2$ & \pt~Parameter (Degree 2) & $\mathcal{U}(0,500)$ & 99.703 \\
    $a_1$ & \pt~Parameter (Degree 1) & $\mathcal{U}(0,500)$ & 146.626 \\
    $a_0$ & \pt~Parameter (Degree 0) & $\mathcal{U}(0,1000)$ & 285.218 \\
    $\log_{10}\left(\Ps\left[\mathrm{bar}\right]\right)$ & Surface Pressure & $\mathcal{U}(-4,3)$ & 0.006 \\
    $\Rpl\,\left[R_\oplus\right]$ & Planet Radius (bulk value) &  $\mathcal{U}(0.5,2)$ & 1.000 \\
    $\Mpl\,\left[M_\oplus\right])$ & Planet Mass (bulk value) & $\mathcal{G}(1.0,0.5)$ & 1.000 \\
    $\log_{10}(\mathrm{N_2})$ & \ce{N2} Mass Fraction & $\mathcal{U}(-15,0)$ & -0.103 \\
    $\log_{10}(\mathrm{O_2})$ & \ce{O2} Mass Fraction & $\mathcal{U}(-15,0)$ & -0.679 \\
    $\log_{10}(\mathrm{H_2O})$ & \ce{H2O} Mass Fraction & $\mathcal{U}(-15,0)$ & -3.000 \\ 
    $\log_{10}(\mathrm{CO_2})$ & \ce{CO2} Mass Fraction & $\mathcal{U}(-15,0)$ & -3.387 \\
    $\log_{10}(\mathrm{CH_4})$ & \ce{CH4} Mass Fraction & $\mathcal{U}(-15,0)$ & -5.770 \\
    $\log_{10}(\mathrm{O_3})$ & \ce{O3} Mass Fraction & $\mathcal{U}(-15,0)$ & -6.523 \\
    $\log_{10}(\mathrm{CO})$ & \ce{CO} Mass Fraction & $\mathcal{U}(-15,0)$ & -6.903 \\
    $\log_{10}(\mathrm{N_2O})$ & \ce{N2O} Mass Fraction & $\mathcal{U}(-15,0)$ & -6.495 \\
    $r_s$ & Surface Reflectance & $\mathcal{U}(0,1)$ & 0.1 \\
    \hline 
\end{tabular}
\tablefoot{$\mathcal{U}(x,y)$ denotes a boxcar prior with a lower threshold $x$ and upper threshold $y$; $\mathcal{G}(\mu,\sigma)$ represents a Gaussian prior with mean $\mu$ and standard deviation $\sigma$).  }

\end{table*}
In this section, we collect some ancillary plots and tables. In Table \ref{table:parameters} we list all the parameters, the priors, and the true values assumed in the retrievals. In Tables \ref{table:opacities}, \ref{table:cia}, \ref{tab:lifesim}, and \ref{tab:luvoirsim} we show details concerning the molecular opacities, the CIA and the Rayleigh opacities used in the forward model, and the simulation settings used in \lifesim~and the PSG simulators.
We report in Tables \ref{apptab:sigmaconstanterrorbar} and \ref{apptab:sigmascaledsnr} the Bayes factor for the two baseline sets of retrievals.
In Figures \ref{appfig:constanterrorbarscorner}, and \ref{appfig:scaledsnrcorner} we show the corner plots and the retrieved estimates of relevant parameters for the two baseline retrieval runs.

\begin{table}[]

\caption{References for the molecular opacities used in the retrievals. } % title of Table
\label{table:opacities} % is used to refer this table in the text
\centering % used for centering table
\setlength\extrarowheight{1pt} 
\begin{tabular}{llll} % centered columns (4 columns)
\hline\hline % inserts double horizontal lines
Species & Line List & Broadening & Line Cutoff\\
\hline
 \ce{O2} &  HITRAN 2020 \emph{(1)}  &  $\gamma_{\mathrm{air}}$&  25 $\mathrm{cm^{-1}}$ \\
 \ce{CO2} &  HITRAN 2020 \emph{(1)}  &  $\gamma_{\mathrm{air}}$&  25 $\mathrm{cm^{-1}}$ \\
\ce{O3} & HITRAN 2020 \emph{(1)}  &  $\gamma_{\mathrm{air}}$&  25 $\mathrm{cm^{-1}}$ \\
 \ce{CH4} & HITRAN 2020 \emph{(1)}  & $\gamma_{\mathrm{air}}$ & 25 $\mathrm{cm^{-1}}$  \\
\ce{CO} &  HITRAN 2020 \emph{(1)}& $\gamma_{\mathrm{air}}$ &  25 $\mathrm{cm^{-1}}$ \\
\ce{H2O} &  HITRAN 2020 \emph{(1)} & $\gamma_{\mathrm{air}}$ &  25 $\mathrm{cm^{-1}}$ \\
\ce{N2O} & HITRAN 2020 \emph{(1)}  & $\gamma_{\mathrm{air}}$ &  25 $\mathrm{cm^{-1}}$ \\\hline

\hline 
\end{tabular}

\tablebib{
\emph{(1)} \cite{GORDON2022107949}
}
\end{table}

\begin{table}[]

\caption{References for the CIA and Rayleigh opacities used in the retrievals. } % title of Table
\label{table:cia} % is used to refer this table in the text
\centering % used for centering table
\setlength\extrarowheight{1pt} 
\begin{tabular}{ll|ll} % centered columns (4 columns)
\hline\hline % inserts double horizontal lines
CIA & References & Rayleigh & References\\
\hline

\ce{N2-N2} & HITRAN \emph{(1)} & \ce{N2} & \emph{(3,4)} \\
 \ce{O2-O2} & HITRAN \emph{(1)} & \ce{O2} & \emph{(3,4)} \\
 \ce{O2-N2} & HITRAN \emph{(1)} & \ce{CO2} & \emph{(5)} \\
 \ce{CO2-CO2} & HITRAN \emph{(1)} & \ce{H2O} & \emph{(6)}\\
 \ce{CH4-CH4} & HITRAN \emph{(1)} & \ce{CH4} & \emph{(5)} \\
 \ce{H2O-H2O} & MT\_CKD \emph{(2)}&  \ce{CO} & \emph{(5)}\\
 \ce{H2O-N2} & MT\_CKD \emph{(2)} &  & \\
\hline 
\end{tabular}
\tablebib{
(1)~\citet{KARMAN2019160}; (2) ~\citet{kofman_absorption_2021}; (3) ~\citet{2014JQSRT.147..171T}; (4) ~\citet{2017JQSRT.189..281T}; (5) ~\citet{2005JQSRT..92..293S}; 
(6)~\citet{1998JPCRD..27..761H}.
}
\end{table}

\begin{table}[]
\caption{Simulation settings used in \lifesim.} 
\label{tab:lifesim} 
\centering 
        \setlength\extrarowheight{1pt} 
\begin{tabular}{l l} 
\hline\hline 
Parameter & \lifesim \\ 
\hline 
Interferometric Baseline (m) & 14.5 \\
 Aperture Diameter (m) & 2~m \\
Wavelength (\textmu m) & 4.0--18.5 \\
 Spectral Resolution & 50 \\
 Reference \SN~\tablefootmark{a} & 10\\
 Reference Wavelength (\textmu m)  \tablefootmark{a}& 11.2\\
 
 Detector quantum efficiency & 0.7\\ % inserting body of the table
 Total instrument throughput & 0.05\\

\hline
\end{tabular}
\tablefoot{See \pI\ and \pII\ for details.~\tablefootmark{a}The \SN~is initially calculated for a set exposure time (1 hr), then scaled to the reference \SN~value at the reference wavelength. }
\end{table}

\begin{table}[]
\caption{Simulation settings used in the PSG ``6-m LUVOIR-B'' simulator for each wavelength bandpass.} 
\label{tab:luvoirsim} 
\centering 
\setlength\extrarowheight{1pt} 
\begin{tabular}{llll} 
\hline\hline 
Parameter & \multicolumn{3}{c}{\hwo} \\ 
\hline 
 Aperture Diameter (m) & \multicolumn{3}{c}{6}  \\
  Number of exposures & \multicolumn{3}{c}{1}\\
  Number of pixels & \multicolumn{3}{c}{10}\\

 Reference \SN~\tablefootmark{a}& \multicolumn{3}{c}{10}\\
 Emissivity of the optics & \multicolumn{3}{c}{0.1} \\
 Temperature of the optics (K) & \multicolumn{3}{c}{270} \\
 Contrast &\multicolumn{3}{c}{$10^{-10}$}\\
 Inner Working Angle&\multicolumn{3}{c}{$\approx4\lambda/D$}\\
 \hline
 & UV&VIS&NIR\\\hline
 Wavelength (\textmu m) & 0.2--0.5 & 0.5--1.0 & 1.0--2.0 \\
  Spectral Resolution & 7 &140 & 70 \\
  Reference Wavelength (\textmu m)~\tablefootmark{a}  & 0.35 & 0.55 & 1.20\\

Average instrument efficiency &  0.05 & 0.15 & 0.2 \\

Read noise ($e^-/px$) &0 & 0& 2.5\\
Dark Current ($e^-/s/px$) & $3\cdot10^{-5}$ & $3\cdot10^{-5}$& $2\cdot10^{-3}$\\
Beam [FWHM] (arcsec) & 0.011 &0.024&0.047\\

\hline
\end{tabular}
\tablefoot{\tablefootmark{a}The \SN~is initially calculated for a set exposure time (1 hr), then scaled to the reference \SN~value at the reference wavelength. }
\end{table}

\begin{table}[]
\caption{Physical parameters used in the noise simulation.} 
\label{tab:physical} 
\centering 
\setlength\extrarowheight{1pt} 
\begin{tabular}{l l} 
\hline\hline 
Parameter & Value \\ 
\hline 
Stellar class & G-type\\
Stellar temperature (K) & 5778\\
Planet-star separation (AU) & 1 \\
 Planet radius (${R_\oplus}$) & 1\\
 Distance to the system (pc) & 10\\
 Exozodi level & 4.5 $\times$ local zodiacal dust \\

\hline
\end{tabular}
\end{table}

\begin{table}[htbp]
\centering
\caption{Bayes factor for each retrieved spectrally active species in the second set of retrievals (simplified noise). Values were obtained by performing a retrieval including and excluding the molecule of interest and then calculating the Bayes factor (eq. \ref{eq:bayes}). }
\setlength\extrarowheight{1pt} 
\begin{tabular}{lccc}
\hline\hline

Parameter & \hwo~& \life~& \hwolife~\\
\hline
 
\ce{O2} &	4.6 &	-0.1 &	5.0\\
\ce{H2O} &	38.3 &	 8.4 &	62.7\\
\ce{CO2} & -0.3 & 16.8 & 31.1\\
\ce{CH4}&	-0.4 &	-0.2 &	0.0\\
\ce{O3}&	0.0 &	1.5 & 2.9\\
\ce{CO}&	0.0	& -0.2 &0.1	\\
\ce{N2O}&	-0.3 &	-0.3 &	-0.1\\

\hline
\end{tabular}
\label{apptab:sigmaconstanterrorbar}
\end{table}

\begin{table}[htbp]
\centering
\caption{Bayes factor for each retrieved spectrally active species in the second set of retrievals (PSG/\lifesim~noise). Values were obtained by performing a retrieval including and excluding the molecule of interest and then calculating the Bayes factor (eq. \ref{eq:bayes}). }\setlength\extrarowheight{1pt} 
\begin{tabular}{lccc}
\hline\hline

Parameter & \hwo~& \life~& \hwolife~\\
\hline
 
\ce{O2} &	3.1  &	0.0 &	3.5\\
\ce{H2O} &48.0 &3.5 & 84.4\\
\ce{CO2} &	-0.1 &	8.9 &	15.4\\
\ce{CH4}&	-0.3 &	-0.1 &	0.0\\
\ce{O3}&	5.7&	1.8 & 9.9\\
\ce{CO}&	-0.2	& -0.1 &-0.1\\
\ce{N2O}&	-0.2 &	-0.1 &	-0.2\\

\hline
\end{tabular}
\label{apptab:sigmascaledsnr}
\end{table}

\end{document}